# Identifying Active Sites of the Water-Gas Shift Reaction over Titania Supported Platinum Catalysts under Uncertainty

Eric A. Walker[1], Donald Mitchell[2], Gabriel A. Terejanu[3,*], Andreas Heyden[1,*]

[1]*Department of Chemical Engineering, University of South Carolina, 301 Main Street, Columbia, South Carolina 29208, USA*

[2]*Department of Chemical Engineering, City College of New York, 160 Covenant Avenue, New York, New York, 10031, USA*

[3]*Department of Computer Science and Engineering, University of South Carolina, 301 Main Street, Columbia, South Carolina, 29208, USA*

*Corresponding authors: email: heyden@cec.sc.edu; TEREJANU@cse.sc.edu
phone: 803-777-5025; 803-777-5872

**ABSTRACT**

A comprehensive uncertainty quantification framework has been developed for integrating computational and experimental kinetic data and to identify active sites and reaction mechanisms in catalysis. Three hypotheses regarding the active site for the water-gas shift reaction on Pt/$TiO_2$ catalysts are tested – Pt(111), an edge interface site, and a corner interface site. Uncertainties associated with DFT calculations and model errors of microkinetic models of the active sites are informed and verified using Bayesian inference and predictive validation. Significant evidence is found for the role of the oxide support in the mechanism. Positive evidence is found in support of the edge interface active site over the corner interface site. For the edge interface site, the CO-promoted redox mechanism is found to be the dominant pathway and only at temperatures above 573 K does the classical redox mechanism contribute significantly to the overall rate. At all reaction conditions, water and surface O-H bond dissociation steps at the Pt/$TiO_2$ interface are the main rate controlling steps.

## 1. INTRODUCTION

Key bottlenecks in the rational design of novel heterogeneous catalysts are our limited ability (*i*) to integrate experimental, kinetic data with computational, first principles models and (*ii*) to identify the relevant active sites on the catalyst. In this paper, a framework for overcoming these bottlenecks is presented and applied to the water-gas shift reaction (WGS: $CO + H_2O \rightleftharpoons CO_2 + H_2$) over Pt catalysts supported on titania. Considering that kinetic data such as the turnover frequency and its temperature and pressure dependence are some of the most important parameters characterizing a heterogeneous catalyst, it is these computational and experimental data that we aim to correlate for the identification of active sites. The WGS is the most widely applied reaction in industry for the generation of hydrogen.[1-10] Currently, hydrogen is produced from natural gas sources through a process involving high pressure steam-reforming.[11] This process produces syngas ($CO + H_2 + CO_2$), whose $CO$ and $H_2$ concentration can be adjusted with the addition of water ($H_2O$) by the WGS. At present, there is disagreement in the literature about the active site of the WGS for Pt catalysts on reducible supports such as $TiO_2$. Some have suggested that the Pt phase is the sole active site, corresponding to terrace active sites studied in this work. This metal-only hypothesis rules out the mechanistic involvement of the support. Grabow et al.[12] and Stamatakis et al.[13,14] have proposed Pt(111) and Pt(211) as the active site, with little effect due to crystal surface structure. It is to be noted though that Grabow et al.[12] arrived at good agreement with experiments only after free energies from DFT were adjusted to the data. In contrast, Schneider et al.[15] found a high surface CO coverage for the WGS in simulations on Pt(111) and Pd(111) leading to low turnover frequencies (TOF $s^{-1}$). Also, for Pt(111) and Pd(111) sites, the reaction orders and apparent activation barrier did not match experiments where Pt and Pd nanoparticles were supported by $\gamma$-$Al_2O_3$, a support that has previously been believed to be not active for the WGS.[15] A number of research groups[16-22] have suggested that most likely the interface of the Pt nanoparticle and the reducible support acts as the active site in most conventionally synthesized catalysts for the WGS. Here, it is still unknown if interface corner or edge sites are the most relevant active sites.

Finally, Stephanopoulos et al.[23,24] have suggested that single Pt atoms are active for the WGS and could be the primary active site at low temperatures. A microkinetic model based on parameters obtained from first principles by Ammal and Heyden[25] confirmed the high activity of atomically dispersed cationic platinum on titania supports, but also suggested that at temperatures above 500 K on most conventionally synthesized Pt catalysts the interface of Pt nanoparticles and the oxide support constitutes the most relevant active site. Previously, Heyden et al.[1,26,27] have reported computational models for various reaction mechanisms of the WGS on corner and edge interface sites for a $Pt_8$ nanoparticle supported on rutile $TiO_2(110)$. They argue that all three-dimensional Pt nanoparticles such as $Pt_8$ on $TiO_2(110)$ behave similarly, considering that the interface, oxygen vacancy formation energy is converged with respect to the number of Pt atoms for $Pt_8$, such that their results remains valid for various titania supported Pt nanoparticles.

For microkinetic models based on parameters obtained from first principles to conclusively identify the active site and reaction mechanism for the WGS over Pt nanoparticles on titania supports in the experiments from various (here three) research groups,[17-19] it is necessary to consider all uncertainties and their correlation in the microkinetic models of the various active sites. Here, we pose this problem of identifying (or more properly eliminating) specific active sites as a Bayesian model selection problem among three hypotheses of active sites investigated by DFT and microkinetic modeling, a Pt(111), interface corner, and interface edge active site model (Figure 1). The uncertainties associated with the DFT calculations, sticking coefficients, and microkinetic models are modeled using probabilities informed by computational DFT data and experimental data of turnover frequency, apparent activation barrier, and reaction orders, i.e., it is assumed that errors related to harmonic transition state theory used to compute elementary rate constants in the microkinetic models are small relative to model errors and errors related to DFT.[17-19] It is noted that Bayesian statistics has grown in popularity recently[28-33] due to the availability of sufficient computational power necessary to solve Bayes' formula; however, in computational catalysis, the Bayesian framework has previously not been used to calibrate microkinetic models and perform model selection as well as identify dominant catalytic cycles under uncertainty.

## 2. METHODS

This section introduces the proposed Bayesian framework for identifying active sites in catalysis (or eliminating specific active sites). The Pt(111) model features reactions occurring on the Pt metal only and the other models feature pathways occurring at a three phase boundary (TPB) of a Pt nanoparticle and a reducible oxide support, $TiO_2$, see Figure 1. Even if uncertainty in results spans orders of magnitude, a probability may be assigned to each model reflecting how well it explains the experimental data. This comparison can provide insight about the active site and reaction mechanism driving a reaction, here the WGS. A first step in model selection is to calibrate each site model, i.e., perform a Bayesian *inverse problem* for each site model. The posterior distribution $p(\theta|D,M)$ corresponding to the parameters of one of the three models, i.e., $M = M_{corner}$, is obtained using Bayes' formula.

$$p(\theta|D,M) = \frac{p(D|\theta,M)p(\theta|M)}{p(D|M)} \quad (1)$$

The parameters $\theta$ (these are *not* surface coverages) are the corrections to all intermediate and transition state relative free energies from DFT, corrections to gas molecule free energies as well as hyperparameters. The posterior joint probability distribution represents the desired estimate of the parameters with quantified uncertainties given the experimental data, $D$, and all prior information for the corresponding model. The prior information is encoded in the prior, $p(\theta|M)$, which contains all the uncertainty settings including correlations and thermodynamics corrections as presented in Walker et al.[26]

Four flavors of DFT are used to generate prior distributions in free energy in this work as suggested for this system by Walker et al.[26] The reason to use four flavors of DFT functionals is to allow the uncertainty to account for several possible approaches for treating the electronic structure of the system. First, generalized gradient approximation (GGA) functionals are used including the Perdew-Burke-Ernzerhof (PBE)[34] and Revised Perdew-Burke-Ernzerhof (RPBE)[35,36] functionals. Both GGA functionals are known to predict quite different adsorption energies particularly for species containing a CO or $CO_2$ backbone.[37] Next, the hybrid Heyd-Scuseria-Ernzerhof (HSE)[38] functional that includes exact exchange was included in the uncertainty quantification (UQ) in addition to the M06L meta-GGA functional that has been optimized against a broad set of experimental data including activation barrier information.[39] We note that our overall procedure is independent of the specific functionals used as long as DFT errors are not significantly underestimated. For adsorption processes, collision theory is used with an uncorrelated sticking coefficient corresponding to a free energy barrier with a mean of 0.075 eV and a standard deviation of 0.075 eV as done in our prior work.[26] Overall gas-phase thermodynamics was corrected to NIST data[40] in an unbiased manner using a Dirichlet[41] probability density function of free energy corrections as was done previously by Walker et al.[26] In this way, the thermodynamics correction is uniformly spread among the four gas molecules as it is unknown for the four functionals which molecular free energy is more accurately described than the other gas species.

## 2.1 Likelihood function and model discrepancy

The likelihood function $p(D|\theta, M)$ provides the likelihood of observing the experimental data $D$ given the particular values of the parameters and the uncertainty in the model and experiment. Each experimental data set $D$ consists of six individual measurements.

$$D = \{TOF, \alpha_{CO}, \alpha_{H_2O}, \alpha_{CO_2}, \alpha_{H_2}, E_{app}\} \tag{2}$$

Here, TOF is the turnover frequency, $\alpha_i$ is the reaction order of carbon monoxide, water, carbon dioxide and hydrogen, respectively, and $E_{app}$ $(eV)$ is apparent activation energy.

The six individual measurements are assumed to be independent given model parameters. This translates into the following factorization of the likelihood function.

$$p(D|\theta, M) = p(TOF|\theta, M)p(\alpha_{CO}|\theta, M)p(\alpha_{H_2O}|\theta, M)p(\alpha_{CO_2}|\theta, M)p(\alpha_{H_2}|\theta, M)p(E_{app}|\theta, M) \tag{3}$$

Each individual likelihood function is defined by the discrepancy between the model simulations, e.g., $E_{app}^*$ and experimental data, e.g., $E_{app}$. This discrepancy is due to unaccounted model errors and unknown experimental errors. Namely, it is assumed that the discrepancy is normally distributed with zero mean and unknown variance, e.g., $\sigma_{E_{app}}^2$.

$$E_{app} = E_{app}^* + \epsilon_{E_{app}} \tag{4}$$

Therefore, the fully expanded likelihood function for the WGS calibration is

$$p(D|\theta,M)=\frac{1}{\sqrt{2\pi\sigma_{TOF}^2}}\exp\left(-\frac{1}{2}\frac{(log_{10}TOF-log_{10}TOF^*)}{\sigma_{TOF}^2}\right)\frac{1}{\sqrt{2\pi\sigma_{\alpha_{CO}}^2}}\exp\left(-\frac{1}{2}\frac{\left(\alpha_{CO}-\alpha_{CO}^*\right)^2}{\sigma_{\alpha_{CO}}^2}\right)\times$$

$$\frac{1}{\sqrt{2\pi\sigma_{\alpha_{H_2O}}^2}}\exp\left(-\frac{1}{2}\frac{\left(\alpha_{H_2O}-\alpha_{H_2O}^*\right)^2}{\sigma_{\alpha_{H_2O}}^2}\right)\frac{1}{\sqrt{2\pi\sigma_{\alpha_{CO_2}}^2}}\exp\left(-\frac{1}{2}\frac{\left(\alpha_{CO_2}-\alpha_{CO_2}^*\right)^2}{\sigma_{\alpha_{CO_2}}^2}\right)\times$$

$$\frac{1}{\sqrt{2\pi\sigma_{\alpha_{H_2}}^2}}\exp\left(-\frac{1}{2}\frac{\left(\alpha_{H_2}-\alpha_{H_2}^*\right)^2}{\sigma_{\alpha_{H_2}}^2}\right)\frac{1}{\sqrt{2\pi\sigma_{E_{app}}^2}}\exp\left(-\frac{1}{2}\frac{\left(E_{app}-E_{app}^*\right)^2}{\sigma_{E_{app}}^2}\right) \quad (5)$$

The hyperparameters $\sigma_{TOF}^2$, $\sigma_{\alpha_{CO}}^2$, $\sigma_{\alpha_{H_2O}}^2$, $\sigma_{CO_2}^2$, $\sigma_{H_2}^2$, $\sigma_{E_{app}}^2$ are calibrated along with DFT and gas molecule corrections. The standard deviations of discrepancies are given prior inverse gamma probability density functions (pdfs), which allows them to extend to infinity, however with most of the probability concentrated around a prior value (see Section I (b) of the supporting information). Note that in all models, there is a constraint on the parameters such that the activation barrier for any elementary step is guaranteed to be non-negative (otherwise transition state theory would not be valid).

When $N$ experimental data sets are available, $\{D\}_{i=1..N}$, (in this case three), it is assumed that they are independent and identically distributed. Independent and identically distributed means that the experiments are not correlated with each other and the experiments have the same uncertainty distribution. The likelihood function is,

$$p(\{D\}_{i=1..N}|\theta,M)=\prod_{i=1}^{N}p(D_i|\theta,M) \quad (6)$$

**2.2 Bayesian model selection**

The marginal likelihood $p(D|M)$, also called evidence in Bayesian model calibration, Eq. (1), acts as both a normalization constant as well as a key quantity in comparing candidate models. It is a natural formulation of Occam's razor, providing an automatic trade-off between goodness-of-fit and model complexity. After model calibration, the evidence in Eq. (1), $p(D|M)$ is used in a second Bayesian inference problem to calculate the posterior probability for all candidate models.

$$p(M|D) = \frac{p(D|M)p(M)}{p(D)} \tag{7}$$

The log-evidence can be written as the difference between the expected log-likelihood of the data and the Kullback-Leibler[42,43] (KL) divergence between posterior and prior pdf of model parameters. The expected log-likelihood quantifies how well the model fits the data, and the KL divergence quantifies model complexity. A large divergence between the posterior and prior pdfs suggests over-fitting of experimental data. Therefore, a complex model is penalized, meaning it might not be selected over a simpler model that does not explain the data as well. This explains its parsimonious model selection property related to Occam's razor. Note, that KL divergence has also been used by Walker et al.[26] to determine the distance of two catalytic cycle TOF ($s^{-1}$) pdf's divergence from the overall TOF ($s^{-1}$). Thus, KL divergence served as a formalization to determining the dominant pathway.

In the absence of information regarding which model is better at describing the catalytic mechanism, the prior model probabilities in Eq. (7) are set to $p(M_{edge}) = p(M_{corner}) = p(M_{terrace}) = \frac{1}{3}$. Note that in this case, the evidence can be used directly to compare the proposed models, and the strength of model comparison can be determined using Bayes' factors. Once both TPB models and the Pt(111) model are calibrated using the same data, the evidences, $p(D|M)$, of each calibration can be divided to produce a Bayes' factor. We implicitly assume here that one and only one active site dominates the observed reaction behavior.

$$B_{edge/corner} = \frac{p(D|M_{edge})}{p(D|M_{corner})} \quad (8)$$

To determine the strength of Bayes' factor in favor of one model against the other, we use Jeffreys scale[44] as shown in Table 1.

The supporting information details further complexities in regard to the order of experimental data points used for the Bayesian inverse, sampling from the posterior probability density, $p(\theta|D, M)$ and approximating the model evidence, $p(D|M)$. Also, all DFT data used for construction of prior distributions for the three active sites are summarized. Finally, lateral interactions for the interface corner and edge active sites are incorporated explicitly, and for the Pt terrace model, they are included with a linear lateral interaction model based on PBE data as described in section IV in the supporting information.

## 3. RESULTS AND DISCUSSION

Each probabilistic model consists of a microkinetic model, a probabilistic discrepancy model to account for errors between model predictions and observations, and a prior distribution over the free energies of intermediates, transition states (SI. I (c), gas molecule corrections and model discrepancy parameters (SI. I(d)) . The results corresponding to the model selection are discussed first, followed by the specific findings for the active site.

**Model selection.** Three datasets (D1[19], D2[17], D3[18] – see SI. II) comprising TOF, apparent activation barrier, and reaction order measurements corresponding to different experimental conditions are used to inform, rank, and validate the proposed probabilistic models. The first experimental dataset is used to constrain the initial DFT-based uncertainty for intermediates and transition states along with the distribution that governs gas molecule corrections and model discrepancies via a Bayesian model calibration (SI. I (a)). This informed distribution becomes the prior distribution in the Bayesian model selection where the second dataset is used to rank the models based on their posterior model probability or evidence in the case of equal prior model probabilities. Finally, the third dataset is used to perform predictive validation to assess the consistency between probabilistic model predictions and experimental data (see SI. III for computational details).

Given that Bayesian model selection is highly sensitive to the prior distribution,[45] all six possible permutations of the three datasets are used to inform, rank and validate to provide robust findings given all available information.[46] Table 2 summarizes the evidence and corresponding Bayes factors with respect to the ranking dataset. In all six cases, the evidence for the terrace site is significantly smaller than for the interface corner and edge sites, which results in "very strong evidence" (see Table 1 - Jeffreys scale[44]) that the terrace site is not the active site among the three. Since the terrace site is the only site which does not include the $TiO_2$ support in the mechanism, a first conclusion may be drawn that the oxide support is mechanistically involved in the WGS.

The evidences for the edge and corner sites are not sufficient to further discriminate between them. The evidences are highly dependent on the datasets used to inform the prior. In Table 2 - Group III of permutations, informing the prior using data D1 and ranking the models on data D2 results in positive evidence for the corner site, however, when informing the prior using D2 and ranking on D1, we find positive evidence for the edge site. Given that Bayesian calibration does not guarantee consistency between model predictions and experimental data, a posterior predictive check is used to solve this ambiguity in model selection.[47] Mahalanobis distance is used as a consistency check to test whether the experimental data set is a possible outcome of the model considering all quantified uncertainties (SI. III).

Even though the Bayes factor indicates that the edge is preferred over the corner site, the consistency check results (the Mahalanobis distance larger than the required threshold) corresponding to Table 2 - Group II of permutations suggest that the datasets D1 and D3 are not sufficient to inform the uncertainty in either of the models and the corresponding results should not be used in model predictions[46] or to discriminate between the two interface sites. In the case of Group III of permutations, the calibrated model corresponding to the corner site has a favorable Bayes factor as compared with the edge, however it does fail the consistency check for both the calibration and validation datasets. The same situation arises in Table 2 - Group I of permutations. Overall, these consistency checks provide positive evidence that the edge site is a better descriptor of the observed catalytic activity given all the available information in this study.

**Edge active site.** Data from two experiments (D1, D2 – Group III) and (D2, D3 – Group I) are used in the Bayesian framework to calibrate the microkinetic model and obtain the posterior predictive distributions for various quantities of interest. Figure 2 depicts the posterior predictive uncertainty for the overall TOF, apparent activation energy (eV) and reaction orders along with the third experimental dataset used for validation purposes (D1 for Group I and D3 for Group III). Compared with the prior predictive uncertainty (based on DFT data only),[26] the posterior uncertainty is reduced while capturing the validation experimental data within the bulk of the probability mass. As with the prior predictive uncertainty, the posterior predictive uncertainty indicates that the CO-promoted redox mechanism is dominant, see Figure 3. It is dominant over the classical redox mechanism in Group III and over the formate mechanism in Group I. This agrees with the free energy pathway being lower for the CO-promoted pathway illustrated in Figure 4. The prior mean corresponding to the edge active site for the dominant CO-promoted redox pathway is slightly higher than the PBE values. The gas molecule corrections provide exact thermodynamics and as a result, no uncertainty is associated with the end of the classical redox pathway shown in Figure 4 (Group III). Also, all free energies are referenced with respect to state S01 such that there is no uncertainty associated with this state. The highest free energy transition state within the CO-promoted redox pathway is the oxygen vacancy formation.

The uncertainties in DFT energies and molecule corrections induce a probability distribution over the degrees of rate control (DRC)[48-52] which are a measure of the rate controlling steps in the reaction network. Figure 5 shows the mean of the degree of rate control corresponding to various transition states. Key rate controlling steps common to Group I and III simulations at reaction conditions corresponding to validation data D1 and D3, respectively, are TS10 ($CO_{Pt}$-$V_{int}$ +$O_{int}$ + $H_2O(g)$ → $CO_{Pt}$-$2OH_{int}$), TS11 ($CO_{Pt}$-$2OH_{int}$ + $O_s$ → $CO_{Pt}$-$OH_{int}$-$O_{int}$-$OH_s$), and TS12 ($CO_{Pt}$-$OH_{int}$-$O_{int}$-$OH_s$ + $O_{int}$ → $CO_{Pt}$-$OH_{int}$-$O_{int}$-$OH_{int}$ + $O_s$). All of these elementary reactions belong to the CO-promoted redox pathway and are water and surface O-H bond dissociations at the Pt/$TiO_2$ interface. In addition, for Group I (low temperature conditions D1), TS08 ($CO_{2(Pt\text{-}int)}$ + CO(g) → $CO_{Pt}$-$CO_{2(int)}$), which is a CO adsorption step on a small coverage site in the CO-promoted redox pathway, becomes partially rate controlling. For Group III (high temperature conditions D3), a surface O-H bond dissociation (TS06: $H_{Pt}$-$OH_{int}$ + $*_{Pt}$ → $2H_{Pt}$-$O_{int}$) and an oxygen vacancy formation step (TS03: $CO_{2(Pt\text{-}int)}$ → $*_{Pt}$-$V_{int}$ + $CO_2(g)$) in the classical redox pathway also become partially rate controlling, illustrating that at temperatures of 573 K both reaction mechanisms, the classical redox and the CO-promoted redox pathway, are operational.

## 4. CONCLUSIONS

Computational catalysis suffers from significant uncertainties in its predictions, primarily due to significant uncertainties in DFT energies. Even when DFT results are combined with microkinetic modeling to simulate experiments from first principles, it is often not possible to conclusively determine when a model (consisting of an active site and reaction mechanism) contains the necessary physics and chemistry to describe experimental, kinetic data. Here, a comprehensive uncertainty quantification framework has been developed for integrating computational and experimental, kinetic catalyst data and to identify active sites and reaction mechanisms in catalysis. This framework was applied to the water-gas shift reaction over Pt catalysts supported on titania. Three actives sites, a Pt(111) terrace model, an edge and a corner interface model, are investigated and the most active site is selected. Four qualitatively different DFT functionals are used to evaluate the prior uncertainty for each site. Using corresponding microkinetic models derived from first principles and experimental kinetic data of TOF, reaction orders and apparent activation barrier, a Bayesian calibration is conducted for each active site model. The evidence for the edge and corner site is significantly larger than for the terrace site, which suggests that the terrace site is not the active site in the experiments. The edge and corner sites are both at the three-phase boundary of the Pt nanoparticle and the $TiO_2$ support; thus, we conclude that the support plays a mechanistic role. Posterior predictive checks are used to discriminate between the edge and the corner site. Consistency between model predictions and experimental data is found in favor of the edge site, which can capture both calibration and validation datasets. Given all available information in this study, we conclude that the edge site is the active site for the WGS in the catalysts studied experimentally when compared with the terrace and interface corner site. Even in the presence of uncertainty, the CO-promoted redox mechanism at the edge active site is found to be the dominant reaction mechanism and only at temperatures above 573 K does the classical redox mechanism contribute also significantly to the overall rate. The prediction of degrees of rate control with quantified uncertainties reveals that at all reaction conditions, water and surface O-H bond dissociation steps at the Pt/$TiO_2$ interface are the main rate controlling steps. At low temperatures of 503 K, a CO adsorption step on a small coverage site in the CO-promoted redox pathway also becomes partially rate controlling while at high temperatures of 573 K an interface $TiO_2$ oxygen vacancy formation step in the classical redox pathway becomes partially rate controlling. Overall, we believe that beyond solving what is the active site for the water-gas shift in the experimental datasets of Pt/$TiO_2$ catalysts, the methodology presented in this work is transferrable to other catalysis challenges where determination of the active site is of critical importance.

## ASSOCIATED CONTENT

### Supporting Information

The Supporting Information is available free of charge on the ACS Publications website at DOI: XXX

Details of proposed Bayesian framework for identifying active sites in catalysis (or eliminating specific active sites) such as the order of data for points for Bayesian inverse, Computational approach, Prior construction for intermediate and transition states, Prior construction for gas molecule corrections and model discrepancy, Experimental data, Model validation via posterior predictive check, Terrance active site model, and Lateral interaction model.

### Notes

The authors declare no competing financial interest.

## ACKNOWLEDGEMENTS

This work was supported by the National Science Foundation under Grant No. CBET-1254352. G.A.T has been supported by NSF DMREF-1534260. D.M. acknowledges funding from NSF EEC-1358931. Furthermore, a portion of this research was performed using XSEDE resources provided by the National Institute for Computational Sciences (NICS), San Diego Supercomputer Center, and Texas advanced Computing Center (TACC) under grant number TG-CTS090100 and at the U.S. Department of Energy facilities located at the National Energy Research Scientific Computing Center (NERSC).

**Table 1**. Jeffreys scale[44] for Bayes factors, $B_{12} = p(D|M_1)/p(D|M_2)$.

| $B_{12}$ | Evidence against $M_2$ |
|---|---|
| 1-3.2 | Not worth more than a bare mention |
| 3.2-10 | Positive |
| 10-100 | Strong |
| >100 | Very Strong |

**Table 2.** Evidences and squared Mahalanobis distances for corner and edge sites and experimental conditions for the three datasets. Positive evidences as per Jeffreys' scale[44] (Table 1) and squared Mahalanobis distances smaller than 12.59 and marked with bold, correspond to consistencies between model predictions and calibration and validation data at 0.05 significance level (see SI. III). Note that the evidences for the terrace site are at least 6 orders of magnitude smaller than evidences of corner and edge for all six possible cases. As a result, the metal-only terrace site is not active. Shaded data points in the table are used to validate model predictions, see also Figure 2.

| Group | Datasets | | Evidence | | | Bayes Factor E/C | Squared Mahalanobis Distance (<=12.59) | | | | | |
|---|---|---|---|---|---|---|---|---|---|---|---|---|
| | | | | | | | D1 | | D2 | | D3 | |
| | Inform | Rank | Corner (C) | Edge (E) | Terrace (T) | | C | E | C | E | C | E |
| I | $D_2$ | $D_3$ | $1.50\times10^{-3}$ | **$5.96\times10^{-3}$** | $5.70\times10^{-12}$ | **3.96** | 23.96 | **2.72** | 29.88 | **12.14** | 21.82 | **10.30** |
| | $D_3$ | $D_2$ | $1.02\times10^{-3}$ | $1.34\times10^{-3}$ | $5.92\times10^{-9}$ | 1.31 | | | | | | |
| II | $D_1$ | $D_3$ | $5.52\times10^{-4}$ | $3.10\times10^{-4}$ | $1.37\times10^{-10}$ | 0.56 | 18.15 | 3.82 | 20.31 | 15.57 | 16.06 | 13.67 |
| | $D_3$ | $D_1$ | $6.71\times10^{-4}$ | $1.90\times10^{-3}$ | $1.50\times10^{-10}$ | 2.83 | | | | | | |
| III | $D_1$ | $D_2$ | **$9.24\times10^{-4}$** | $2.19\times10^{-4}$ | $1.47\times10^{-36}$ | **0.24** | 16.58 | **3.18** | 18.17 | **8.69** | 13.41 | **7.72** |
| | $D_2$ | $D_1$ | $1.66\times10^{-3}$ | **$5.99\times10^{-3}$** | $1.67\times10^{-33}$ | **3.61** | | | | | | |

| | |
|---|---|
| $D1$[19] | $P_{CO}$ = 0.07 atm, $P_{H_2O}$ = 0.22 atm, $P_{CO_2}$ = 0.09 atm, $P_{H_2}$ = 0.37 atm, T = 503 K |
| $D2$[17] | $P_{CO}$ = 0.03 atm, $P_{H_2O}$ = 0.10 atm, $P_{CO_2}$ = 0.06 atm, $P_{H_2}$ = 0.20 atm, T = 523 K |
| $D3$[18] | $P_{CO}$ = 0.10 atm, $P_{H_2O}$ = 0.20 atm, $P_{CO_2}$ = 0.10 atm, $P_{H_2}$ = 0.40 atm, T = 573 K |

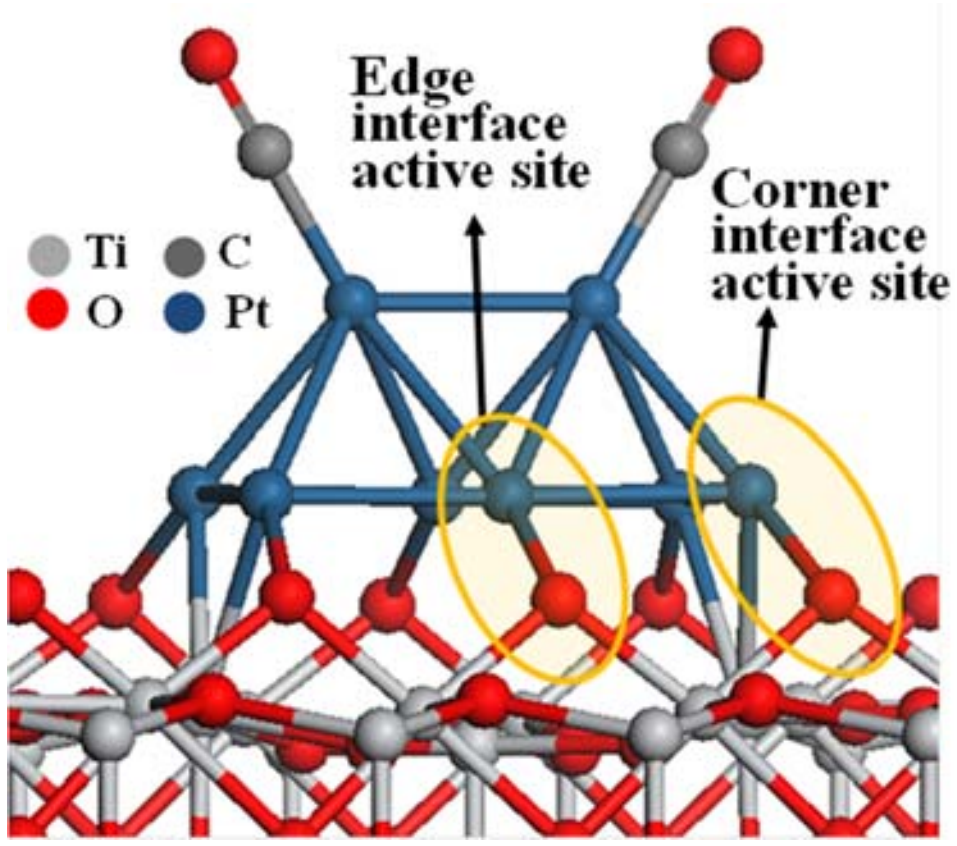

**Figure 1.** Pt/$TiO_2$ catalyst model with highlighted edge and corner interface sites used to study the WGS reaction mechanism.

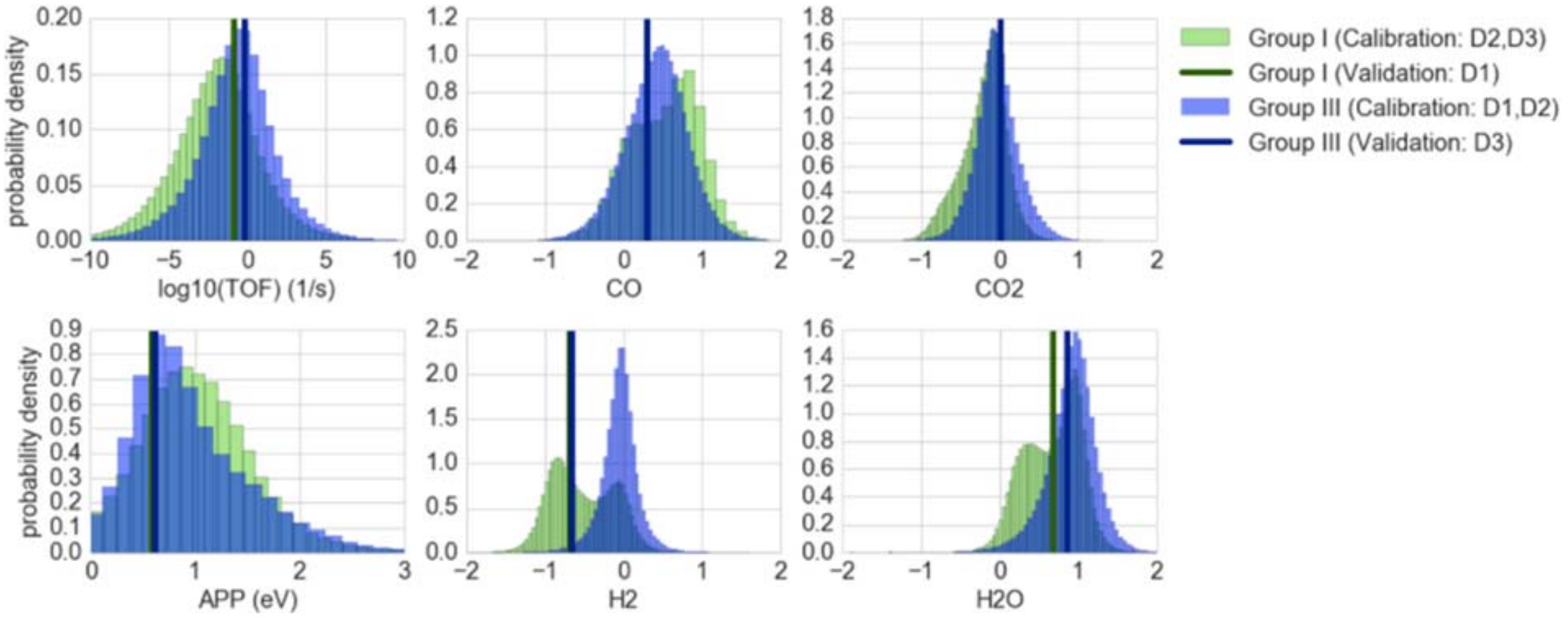

**Figure 2.** Posterior predictive distributions for turnover frequency (TOF), apparent activation barrier (APP), and the four reaction orders CO, $CO_2$, $H_2$, and $H_2O$ for the edge interface active site at experimental conditions corresponding to validation data of Group I (D1) and Group III (D3), see Table 2. Corresponding validation data is also shown. Validation data for the CO and $CO_2$ reaction order overlap.

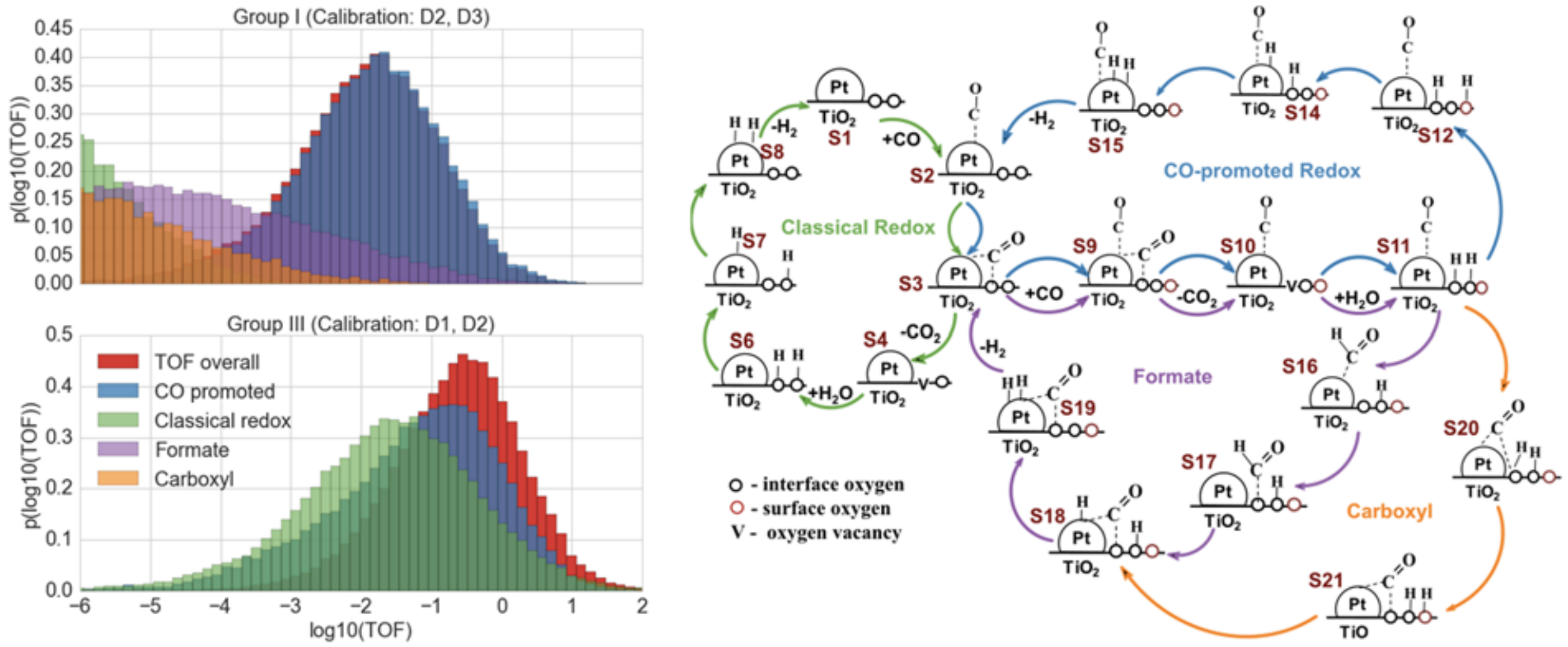

**Figure 3.** [Left] Overall TOF ($s^{-1}$) and TOF ($s^{-1}$) of individual catalytic cycles for the edge interface active site for both Group I and Group III, see Table 2, at reaction conditions of D1 and D3, respectively. The dominant catalytic cycle is the CO-promoted in both cases followed by the formate mechanism in case of Group I and the classical redox mechanism for Group III. [Right] Reaction network of possible WGS reaction steps corresponding to the edge site.

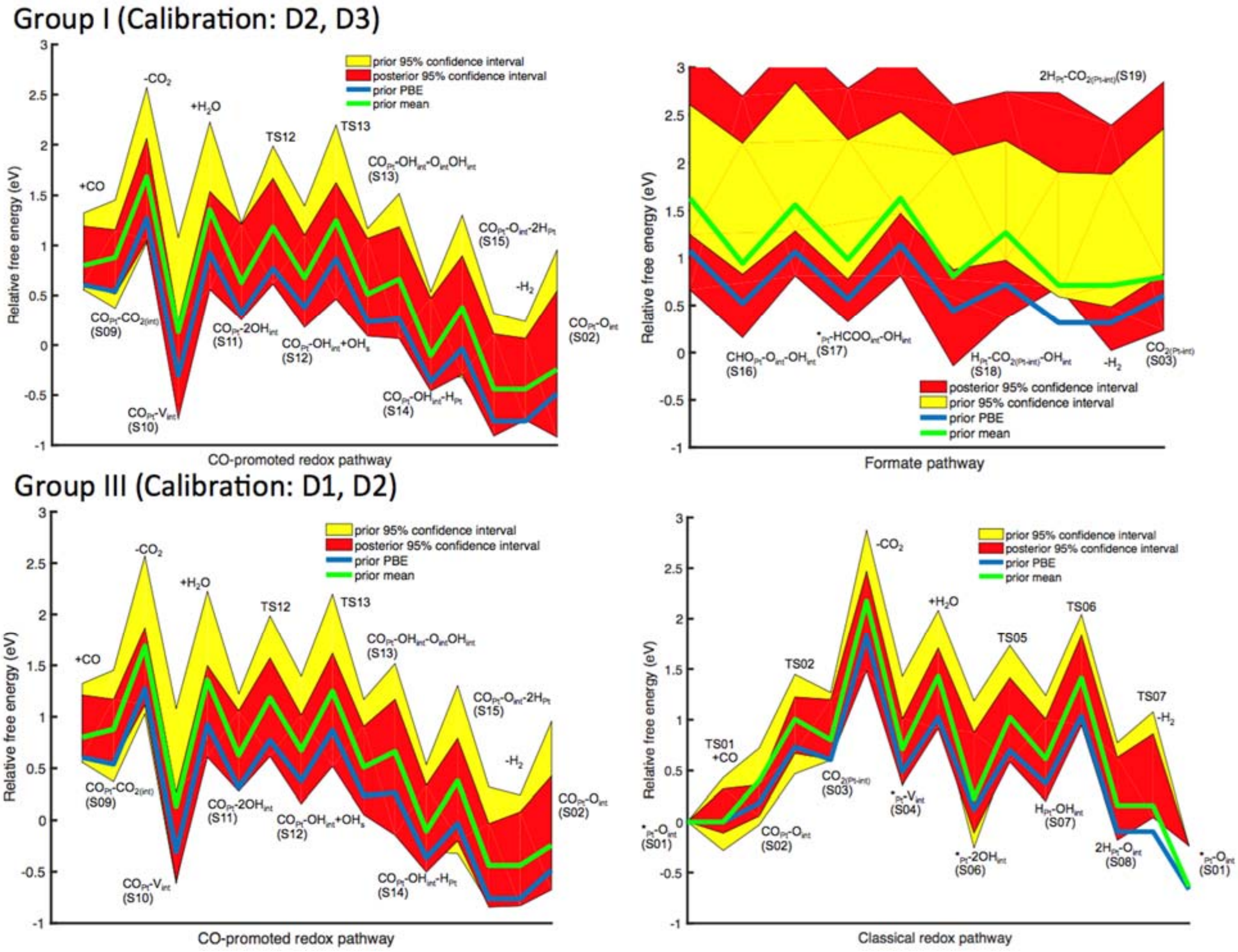

**Figure 4.** Prior and posterior uncertainty for the dominant free energy pathway (CO-promoted redox mechanism) for the edge interface active site for both Group I evaluated at data set D1 and Group III evaluated at data set D3 and the runner-up free energy pathway (Group I – formate mechanism and Group III – classical redox mechanism). Also shown are the free energies obtained from PBE and the prior mean of the four functionals *without* thermodynamics correction. All free energies are referenced with respect to state S1 shown in Figure 1 with 2CO and $H_2O$ gas molecules. D1[19] conditions are $P_{CO} = 0.07\ atm, P_{H_2O} = 0.22\ atm, P_{CO_2} = 0.09\ atm, P_{H_2} = 0.37\ atm, T = 503\ K$. D3[18] conditions are $P_{CO} = 0.10\ atm, P_{H_2O} = 0.20\ atm, P_{CO_2} = 0.10\ atm, P_{H_2} = 0.40\ atm, T = 573\ K$.

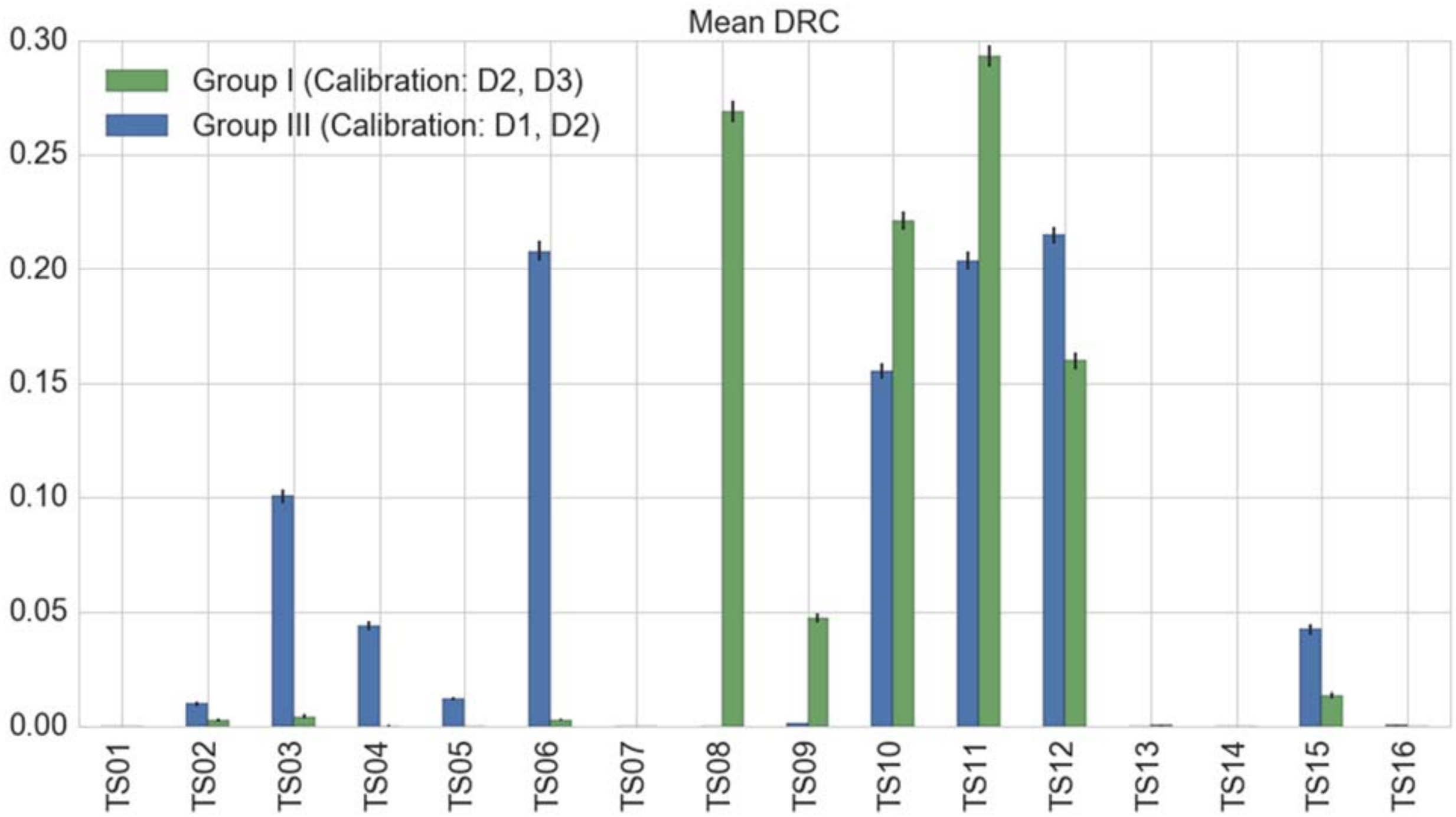

**Figure 5.** Mean degree of rate control for the edge interface active site for both Group I and Group III evaluated at conditions corresponding to validation data D1 and D3, respectively. Table S1 in the supporting information describes the physical meaning of all transition state (TS) labels. Error bars correspond to the standard errors of the mean.

**Supporting information:**

# Identifying Active Sites of the Water-Gas Shift over Titania Supported Platinum Catalysts under Uncertainty

Eric A. Walker[1], Donald Mitchell[2], Gabriel A. Terejanu[3,*], Andreas Heyden[1,*]

[1]*Department of Chemical Engineering, University of South Carolina, 301 Main Street, Columbia, South Carolina 29208, USA*

[2]*Department of Chemical Engineering, City College of New York, 160 Covenant Avenue, New York, New York, 10031, USA*

[3]*Department of Computer Science and Engineering, University of South Carolina, 301 Main Street, Columbia, South Carolina, 29208, USA*

## Contents

## I. Bayesian statistics

### I (a) Order of data points for Bayesian inverse

Due to the dependence of model evidence on the prior distribution, especially for parts of the prior that are weakly informed such as model discrepancies, combinations of the three experimental data points are used in different orders and the evidences are listed in Table 2 of the main article. One experimental dataset, e.g., $D1$, is used to further constrain the initial prior $p\ (\theta|M)$ corresponding to the correction of intermediates and transition states, gas molecule corrections and model error which at first are weakly informed, see Eq. (S1).

$$p(\theta|D_1, M) = \frac{p(D_1|\theta, M)p(\theta|M)}{p(D_1|M)} \tag{S1}$$

The posterior distribution, $p(\theta|D_1, M)$ becomes the prior distribution in a sequential Bayesian inference, where a second experimental data point is used to compute the evidence for Bayesian model selection using the prior from the first experimental data point. As a result, we use the evidence $p(D_2|D_1, M)$ to compare the three models corresponding to the candidate sites. This evidence calculation may be achieved using sequential Bayes rule, see Eq. (S2).

$$p(\theta|D_2, D_1, M) = \frac{p(D_2|\theta_1, D_1, M)p(\theta|D_1, M)}{p(D_2|D_1, M)} \tag{S2}$$

Computationally, the sampling approach used to obtain the posterior distribution and calculate the model evidence is not appropriate to solve sequential Bayesian inverse problems such as Eq. (S2), see the next section. This is due to the fact that the posterior obtained in Eq. (S1) cannot be accurately evaluated in Eq. (S2) given its representation based on samples. The alternative is to obtain the evidence $p(D_2|D_1, M)$ using two separate batch Bayesian inverse problems. The first Bayesian inverse problem is given by Eq. (S1), where $p(D_1|M)$, the evidence with respect to $D_1$ is obtained. The second Bayesian inverse problem is given by Eq. (S3), where the desired posterior distribution $p(\theta|D_2, D_1, M)$ is obtained as in Eq. (S2), and where $p(D_2, D_1|M)$, the evidence with respect to $D_1$ and $D_2$ is obtained.

$$p(\theta|D_2, D_1, M) = \frac{p(D_2, D_1|\theta, M)p(\theta|M)}{p(D_2, D_1|M)} \tag{S3}$$

Finally, the desired evidence $p(D_2|D_1, M)$ corresponding to the constrained prior distribution, is obtained as the ratio between the evidences of the two batch Bayesian inverse problems as follows.

$$p(D_2|D_1, M) = \frac{p(D_2, D_1|M)}{p(D_1|M)} \tag{S4}$$

## I (b) Computational approach

In general, sampling from the posterior probability density, $p(\theta|D, M)$ and approximating the model evidence, $p(D|M)$, is not a trivial task. Here, we are using the multilevel sampling algorithm in the statistical library QUESO.[1,2] The multilevel algorithm reduces two potential drawbacks of MCMC (Markov chain Monte Carlo) algorithms. First, the MCMC may take too small steps and either not arrive at the high-probability region of the parameter space or the chain may be inside the high probability region but not sample all of it. Second, the steps may be too large that they skip over the high probability region entirely. A sequence of intermediate distributions are sampled on the way to the final target distribution. The first sampled pdf is flattened and becomes more like the prior pdf. This is achieved by the following factorization of the likelihood.

$$p(\theta|D, M) = \frac{\prod_{j=1}^{L} p(D|\theta, M)^{\alpha_j} p(\theta|M)}{p(D|M)} \tag{S5}$$

$$\sum_{j=1}^{L} \alpha_j = 1 \tag{S6}$$

The overall log evidence is the sum of the log evidences at each level of the multilevel sampling where $j$ is the level and $L$ is the total number of levels.[2-4]

## I (c) Prior construction for intermediate and transition states

Previously, Walker et al.[5] used four DFT functionals to obtain a prior (before Bayesian inverse) uncertainty. The same four functionals which were calculated for the edge active site[5] are also used for the corner active site[6] and the Pt(111) active site. Supplementary Table 1 lists the relative free energies as calculated by the four DFT functionals for the edge active site and is a reproduction from Walker, et al.[5] Supplementary Tables 2-3 list the relative free energies as calculated by the four DFT functionals for Pt(111) and corner, respectively.

Supplementary Table 1. Relative free energies as calculated by four DFT functionals[7-12] for the edge active site. Table 1(a) reproduced from Walker et al.[5]

(a) 503 K

| Label | Intermediate or Transition State | G (eV) | | | |
|---|---|---|---|---|---|
| | | PBE | RPBE | HSE | M06L |
| S01 | $*_{Pt}$-$O_{int}$ + 2CO(g) + $H_2O$(g) | 0.000 | 0.000 | 0.000 | 0.000 |
| TS01 | $*_{Pt}$-$O_{int}$ + 2CO(g) + $H_2O$(g) → $CO_{Pt}$-$O_{int}$ + CO(g) + $H_2O$(g) | 0.000 | 0.000 | 0.000 | 0.000 |
| S02 | $CO_{Pt}$-$O_{int}$ + CO(g) + $H_2O$(g) | -0.060 | 0.031 | 0.123 | 0.526 |
| TS02 | $CO_{Pt}$-$O_{int}$ + CO(g) + $H_2O$(g) → $CO_{2(Pt\text{-}int)}$ + CO(g) + $H_2O$(g) | 0.580 | 0.798 | 0.787 | 1.246 |
| S03 | $CO_{2(Pt\text{-}int)}$ + CO(g) + $H_2O$(g) | 0.429 | 0.716 | 0.572 | 0.781 |
| TS03 | $CO_{2(Pt\text{-}int)}$ + CO(g) + $H_2O$(g) → $*_{Pt}$-$V_{int}$ + $CO_2$(g) + CO(g) + $H_2O$(g) | 1.245 | 1.392 | 2.219 | 1.511 |
| S04 | $*_{Pt}$-$V_{int}$ + CO(g) + $H_2O$(g) + $CO_2$(g) | 0.555 | 0.474 | 0.934 | 1.140 |
| TS04 | $*_{Pt}$-$V_{int}$ + CO(g) + $H_2O$(g) + $CO_2$(g) → $*_{Pt}$-$2OH_{int}$ + CO(g) + $CO_2$(g) | 0.966 | 1.209 | 1.990 | 1.341 |
| S06 | $*_{Pt}$-$2OH_{int}$ + CO(g) + $CO_2$(g) | 0.024 | 0.365 | 0.570 | -0.450 |
| TS05 | $*_{Pt}$-$2OH_{int}$ + CO(g) + $CO_2$(g) → $H_{Pt}$-$OH_{int}$ + $O_{int}$ + CO(g) + $CO_2$(g) | 0.631 | 0.839 | 1.081 | 1.283 |
| S07 | $H_{Pt}$-$OH_{int}$ + $*_{Pt}$ + CO(g) + $CO_2$(g) | 0.289 | 0.422 | 0.651 | 0.757 |
| **TS06** | $H_{Pt}$-$OH_{int}$ + $*_{Pt}$ + CO(g) + $CO_2$(g) → $2H_{Pt}$-$O_{int}$ + CO(g) + $CO_2$(g) | 0.870 | 1.258 | 1.687 | 1.173 |
| **S08** | $2H_{Pt}$-$O_{int}$ + CO(g) + $CO_2$(g) | -0.185 | 0.157 | 0.221 | 0.087 |
| **TS07** | $2H_{Pt}$-$O_{int}$ + CO(g) + $CO_2$(g) → $*_{Pt}$-$O_{int}$ +$*_{Pt}$ +$H_2$(g) + CO(g) + $CO_2$(g) | -0.185 | 0.157 | 0.221 | 0.087 |
| **S01** | $*_{Pt}$-$O_{int}$ +$*_{Pt}$ +$H_2$(g) + CO(g) + $CO_2$(g) | -0.683 | -0.638 | -0.484 | -0.819 |
| **TS08** | $CO_{2(Pt\text{-}int)}$ + CO(g) + $H_2O$(g) → $CO_{Pt}$-$CO_{2(int)}$ + $H_2O$(g) | 0.860 | 1.147 | 1.003 | 1.212 |
| **S09** | $CO_{Pt}$-$CO_{2(int)}$ + $H_2O$(g) | 0.047 | 0.287 | 0.386 | 0.825 |
| **TS09** | $CO_{Pt}$-$CO_{2(int)}$ + $H_2O$(g) → $CO_{Pt}$-$V_{int}$ + $CO_2$(g) + $H_2O$(g) | 0.457 | 0.508 | 1.036 | 1.491 |
| **S10** | $CO_{Pt}$-$V_{int}$ +$O_{int}$ + $H_2O$(g) + $CO_2$(g) | -0.471 | -0.429 | 0.441 | 0.331 |
| **TS10** | $CO_{Pt}$-$V_{int}$ +$O_{int}$ $H_2O$(g) + $CO_2$(g) → $CO_{Pt}$-$2OH_{int}$ + $CO_2$(g) | 0.607 | 0.733 | 1.179 | 1.641 |
| **S11** | $CO_{Pt}$-$2OH_{int}$ + $O_s$ + $CO_2$(g) | -0.037 | 0.324 | 0.608 | 0.221 |
| **TS11** | $CO_{Pt}$-$2OH_{int}$ + $O_s$ + $CO_2$(g) → $CO_{Pt}$-$OH_{int}$-$O_{int}$-$OH_s$ + $CO_2$(g) | 0.457 | 0.797 | 0.892 | 1.337 |
| **S12** | $CO_{Pt}$-$OH_{int}$-$O_{int}$-$OH_s$ + $O_{int}$ + $CO_2$ (g) | 0.051 | 0.243 | 0.330 | 0.766 |
| **TS12** | $CO_{Pt}$-$OH_{int}$-$O_{int}$-$OH_s$ + $O_{int}$ + $CO_2$ (g) → $CO_{Pt}$-$OH_{int}$-$O_{int}$-$OH_{int}$ + $O_s$ + $CO_2$ (g) | 0.591 | 0.671 | 0.980 | 1.642 |
| **S13** | $CO_{Pt}$-$OH_{int}$-$O_{int}$-$OH_{int}$ + $*_{Pt}$ + $CO_2$ (g) | -0.064 | 0.217 | 0.477 | 0.214 |
| **TS13** | $CO_{Pt}$-$OH_{int}$-$O_{int}$-$OH_{int}$ + $*_{Pt}$ + $CO_2$ (g)→ $CO_{Pt}$-$OH_{int}$-$O_{int}$-$H_{Pt}$ + $O_{int}$ + $CO_2$ (g) | -0.041 | 0.094 | 0.345 | 1.041 |
| **S14** | $CO_{Pt}$-$OH_{int}$-$O_{int}$-$H_{Pt}$ + $*_{Pt}$ + $CO_2$ (g) | -0.659 | -0.392 | -0.469 | -0.076 |
| **TS14** | $CO_{Pt}$-$OH_{int}$-$O_{int}$-$H_{Pt}$ + $*_{Pt}$ + $CO_2$ (g)→ $CO_{Pt}$-$O_{int}$-$2H_{Pt}$ + $O_{int}$ + $CO_2$ (g) | -0.359 | -0.177 | -0.029 | 0.781 |
| **S15** | $CO_{Pt}$-$O_{int}$-$2H_{Pt}$ + $CO_2$ (g) | -1.054 | -0.776 | -0.905 | -0.191 |
| **TS15** | $CO_{Pt}$-$O_{int}$-$2H_{Pt}$ + $CO_2$ (g)→ $CO_{Pt}$-$O_{int}$ + $2*_{Pt}$ + $CO_2$ (g)+ $H_2$ (g) | -1.054 | -0.776 | -0.905 | -0.191 |
| **S02** | $CO_{Pt}$-$O_{int}$ + $2*_{Pt}$ + $CO_2$ (g)+ $H_2$ (g) | -0.743 | -0.607 | -0.361 | -0.292 |
| **TS16** | $CO_{Pt}$-$2OH_{int}$ + CO(g) + $H_2O$(g) →$CHO_{Pt}$-$O_{int}$-$OH_{int}$+ CO(g) + $H_2O$(g) | 0.776 | 1.149 | 1.623 | 1.772 |
| **S16** | $CHO_{Pt}$-$O_{int}$-$OH_{int}$+ CO(g) + $H_2O$(g) | 0.208 | 0.552 | 0.759 | 0.999 |
| **TS17** | $CHO_{Pt}$-$O_{int}$-$OH_{int}$+ CO(g) + $H_2O$(g) → $*_{Pt}$ –$HCOO_{int}$-$OH_{int}$+ CO(g) + $H_2O$(g) | 0.823 | 1.420 | 1.505 | 1.526 |
| **S17** | $*_{Pt}$ –$HCOO_{int}$-$OH_{int}$+ CO(g) + $H_2O$(g) | 0.265 | 0.909 | 1.226 | 0.336 |
| **TS18** | $*_{Pt}$ –$HCOO_{int}$-$OH_{int}$ + $*_{Pt}$ + CO(g) + $H_2O$(g) → $H_{Pt}$-$CO_{2(Pt\text{-}int)}$-$OH_{int}$+ CO(g) + $H_2O$(g) | 0.900 | 1.454 | 1.886 | 1.326 |
| **S18** | $H_{Pt}$-$CO_{2(Pt\text{-}int)}$-$OH_{int}$+ CO(g) + $H_2O$(g) | 0.181 | 0.679 | 0.423 | 0.901 |
| **TS19** | $H_{Pt}$-$CO_{2(Pt\text{-}int)}$-$OH_{int}$+ $*_{Pt}$ + CO(g) + $H_2O$(g) →2$H_{Pt}$-$CO_{2(Pt\text{-}int)}$+$O_{int}$+ CO(g) + $H_2O$(g) | 0.467 | 0.900 | 0.917 | 1.768 |
| **S19** | 2$H_{Pt}$-$CO_{2(Pt\text{-}int)}$+$O_{int}$+ CO(g) + $H_2O$(g) | 0.054 | 0.620 | 0.374 | 0.747 |
| **TS20** | 2$H_{Pt}$-$CO_{2(Pt\text{-}int)}$ + CO(g) + $H_2O$(g) → $CO_{2(Pt\text{-}int)}$ + 2$*_{Pt}$ + $H_2$(g) + CO(g) + $H_2O$(g) | 0.054 | 0.620 | 0.374 | 0.747 |
| **S03** | $CO_{2(Pt\text{-}int)}$ + 2$*_{Pt}$ + $H_2$(g) + CO(g) + $H_2O$(g) | -0.253 | 0.081 | 0.089 | -0.035 |
| **TS21** | $CO_{Pt}$-2$OH_{int}$ + CO(g) + $H_2O$(g) → $COOH_{(Pt\text{-}int)}$-$OH_{int}$+ CO(g) + $H_2O$(g) | 0.727 | 1.243 | 1.724 | 1.050 |
| **S20** | $COOH_{(Pt\text{-}int)}$-$OH_{int}$+ CO(g) + $H_2O$(g) | 0.687 | 1.219 | 1.624 | 1.041 |
| **TS22** | $COOH_{(Pt\text{-}int)}$-$OH_{int}$+ CO(g) + $H_2O$(g) → $CO_{2(Pt\text{-}int)}$-$OH_s$-$OH_{int}$+ CO(g) + $H_2O$(g) | 0.847 | 1.465 | 1.905 | 1.570 |
| **S21** | $CO_{2(Pt\text{-}int)}$-$OH_s$-$OH_{int}$+ CO(g) + $H_2O$(g) | 0.673 | 1.276 | 1.632 | 1.033 |
| **TS23** | $CO_{2(Pt\text{-}int)}$-$OH_s$-$OH_{int}$+ $O_{int}$ + CO(g) + $H_2O$(g) → $CO_{2(Pt\text{-}int)}$-$OH_{int}$-$OH_{int}$+ $O_s$ + CO(g) + $H_2O$(g) | 1.074 | 1.682 | 2.149 | 1.691 |
| **S22** | $CO_{2(Pt\text{-}int)}$-$OH_{int}$-$OH_{int}$+ $O_s$ + CO(g) + $H_2O$(g) | 0.260 | 0.859 | 1.089 | 0.380 |
| **TS24** | $CO_{2(Pt\text{-}int)}$-$OH_{int}$-$OH_{int}$+ $*_{Pt}$ + CO(g) + $H_2O$(g) → $H_{Pt}$-$CO_{2(Pt\text{-}int)}$-$OH_{int}$+$O_{int}$+ CO(g) + $H_2O$(g) | 0.272 | 0.883 | 1.274 | 1.355 |
| **S18** | $H_{Pt}$-$CO_{2(Pt\text{-}int)}$-$OH_{int}$+$O_{int}$+ CO(g) + $H_2O$(g) | 0.181 | 0.678 | 0.422 | 0.900 |

(b) 523 K

| Label | Intermediate or Transition State | G (eV) | | | |
|---|---|---|---|---|---|
| | | PBE | RPBE | HSE | M06L |
| **S01** | $*_{Pt}$-$O_{int}$ + 2CO(g) + $H_2O$(g) | 0.000 | 0.000 | 0.000 | 0.000 |
| **TS01** | $*_{Pt}$-$O_{int}$ + 2CO(g) + $H_2O$(g)→ $CO_{Pt}$-$O_{int}$ + CO(g) + $H_2O$(g) | 0.000 | 0.000 | 0.000 | 0.000 |
| **S02** | $CO_{Pt}$-$O_{int}$ + CO(g) + $H_2O$(g) | 0.182 | 0.273 | 0.365 | 0.768 |
| **TS02** | $CO_{Pt}$-$O_{int}$ + CO(g) + $H_2O$(g) → $CO_{2(Pt-int)}$ + CO(g) + $H_2O$(g) | 0.733 | 0.951 | 0.940 | 1.399 |
| **S03** | $CO_{2(Pt-int)}$ + CO(g) + $H_2O$(g) | 0.605 | 0.892 | 0.748 | 0.957 |
| **TS03** | $CO_{2(Pt-int)}$ + CO(g) + $H_2O$(g) → $*_{Pt}$-$V_{int}$ + $CO_2$(g) + CO(g) + $H_2O$(g) | 1.829 | 1.976 | 2.803 | 2.095 |
| **S04** | $*_{Pt}$-$V_{int}$ + CO(g) + $H_2O$(g) + $CO_2$(g) | 0.489 | 0.408 | 0.868 | 1.074 |
| **TS04** | $*_{Pt}$-$V_{int}$ + CO(g) + $H_2O$(g) + $CO_2$(g)→ $*_{Pt}$-$2OH_{int}$ + CO(g) + $CO_2$(g) | 1.020 | 1.263 | 2.044 | 1.395 |
| **S06** | $*_{Pt}$-$2OH_{int}$ + CO(g) + $CO_2$(g) | 0.121 | 0.462 | 0.667 | -0.353 |
| **TS05** | $*_{Pt}$-$2OH_{int}$ + CO(g) + $CO_2$(g) → $H_{Pt}$-$OH_{int}$ + $O_{int}$ + CO(g) + $CO_2$(g) | 0.697 | 0.905 | 1.147 | 1.349 |
| **S07** | $H_{Pt}$-$OH_{int}$ + $*_{Pt}$ + CO(g) + $CO_2$(g) | 0.374 | 0.507 | 0.736 | 0.842 |
| **TS06** | $H_{Pt}$-$OH_{int}$ + $*_{Pt}$ + CO(g) + $CO_2$(g) → $2H_{Pt}$-$O_{int}$ + CO(g) + $CO_2$(g) | 1.036 | 1.424 | 1.853 | 1.339 |
| **S08** | $2H_{Pt}$-$O_{int}$ + CO(g) + $CO_2$(g) | -0.097 | 0.245 | 0.309 | 0.175 |
| **TS07** | $2H_{Pt}$-$O_{int}$ + CO(g) + $CO_2$(g) → $*_{Pt}$-$O_{int}$ +$*_{Pt}$ +$H_2$(g) + CO(g) + $CO_2$(g) | -0.097 | 0.245 | 0.309 | 0.175 |
| **S01** | $*_{Pt}$-$O_{int}$ +$*_{Pt}$ +$H_2$(g) + CO(g) + $CO_2$(g) | -0.667 | -0.622 | -0.468 | -0.803 |
| **TS08** | $CO_{2(Pt-int)}$ + CO(g) + $H_2O$(g) → $CO_{Pt}$-$CO_{2(int)}$ + $H_2O$(g) | 0.605 | 0.892 | 0.748 | 0.957 |
| **S09** | $CO_{Pt}$-$CO_{2(int)}$ + $H_2O$(g) | 0.540 | 0.780 | 0.879 | 1.318 |
| **TS09** | $CO_{Pt}$-$CO_{2(int)}$ + $H_2O$(g) → $CO_{Pt}$-$V_{int}$ + $CO_2$(g) + $H_2O$(g) | 1.272 | 1.323 | 1.851 | 2.306 |
| **S10** | $CO_{Pt}$-$V_{int}$ +$O_{int}$ + $H_2O$(g) + $CO_2$(g) | -0.305 | -0.263 | 0.607 | 0.497 |
| **TS10** | $CO_{Pt}$-$V_{int}$ +$O_{int}$ $H_2O$(g) + $CO_2$(g) → $CO_{Pt}$-$2OH_{int}$ + $CO_2$(g) | 0.924 | 1.050 | 1.496 | 1.958 |
| **S11** | $CO_{Pt}$-$2OH_{int}$ + $O_s$ + $CO_2$(g) | 0.313 | 0.673 | 0.957 | 0.570 |
| **TS11** | $CO_{Pt}$-$2OH_{int}$ + $O_s$ + $CO_2$(g) → $CO_{Pt}$-$OH_{int}$-$O_{int}$-$OH_s$ + $CO_2$(g) | 0.771 | 1.111 | 1.206 | 1.651 |
| **S12** | $CO_{Pt}$-$OH_{int}$-$O_{int}$-$OH_s$ + $O_{int}$ + $CO_2$ (g) | 0.379 | 0.571 | 0.658 | 1.094 |
| **TS12** | $CO_{Pt}$-$OH_{int}$-$O_{int}$-$OH_s$ + $O_{int}$ + $CO_2$ (g) → $CO_{Pt}$-$OH_{int}$-$O_{int}$-$OH_{int}$ + $O_s$ + $CO_2$ (g) | 0.869 | 0.949 | 1.258 | 1.920 |
| **S13** | $CO_{Pt}$-$OH_{int}$-$O_{int}$-$OH_{int}$ + $*_{Pt}$ + $CO_2$ (g) | 0.236 | 0.517 | 0.777 | 0.514 |
| **TS13** | $CO_{Pt}$-$OH_{int}$-$O_{int}$-$OH_{int}$ + $*_{Pt}$ + $CO_2$ (g)→ $CO_{Pt}$-$OH_{int}$-$O_{int}$-$H_{Pt}$ + $O_{int}$ + $CO_2$ (g) | 0.262 | 0.397 | 0.648 | 1.344 |
| **S14** | $CO_{Pt}$-$OH_{int}$-$O_{int}$-$H_{Pt}$ + $*_{Pt}$ + $CO_2$ (g) | -0.364 | -0.097 | -0.174 | 0.219 |
| **TS14** | $CO_{Pt}$-$OH_{int}$-$O_{int}$-$H_{Pt}$ + $*_{Pt}$ + $CO_2$ (g)→ $CO_{Pt}$-$O_{int}$-2$H_{Pt}$ + $O_{int}$ + $CO_2$ (g) | -0.034 | 0.148 | 0.296 | 1.106 |
| **S15** | $CO_{Pt}$-$O_{int}$-2$H_{Pt}$ + $CO_2$ (g) | -0.761 | -0.483 | -0.612 | 0.102 |
| **TS15** | $CO_{Pt}$-$O_{int}$-2$H_{Pt}$ + $CO_2$ (g)→ $CO_{Pt}$-$O_{int}$ + 2$*_{Pt}$ + $CO_2$ (g)+ $H_2$ (g) | -0.761 | -0.483 | -0.612 | 0.102 |
| **S02** | $CO_{Pt}$-$O_{int}$ + 2$*_{Pt}$ + $CO_2$ (g)+ $H_2$ (g) | -0.485 | -0.349 | -0.103 | -0.034 |
| **TS16** | $CO_{Pt}$-2$OH_{int}$ + CO(g) + $H_2O$(g) →$CHO_{Pt}$-$O_{int}$-$OH_{int}$+ CO(g) + $H_2O$(g) | 1.077 | 1.450 | 1.924 | 2.073 |
| **S16** | $CHO_{Pt}$-$O_{int}$-$OH_{int}$+ CO(g) + $H_2O$(g) | 0.523 | 0.867 | 1.074 | 1.314 |
| **TS17** | $CHO_{Pt}$-$O_{int}$-$OH_{int}$+ CO(g) + $H_2O$(g) →$*_{Pt}$–$HCOO_{int}$-$OH_{int}$+ CO(g) + $H_2O$(g) | 1.064 | 1.661 | 1.746 | 1.767 |
| **S17** | $*_{Pt}$–$HCOO_{int}$-$OH_{int}$+ CO(g) + $H_2O$(g) | 0.567 | 1.211 | 1.528 | 0.638 |
| **TS18** | $*_{Pt}$–$HCOO_{int}$-$OH_{int}$ + $*_{Pt}$ + CO(g) + $H_2O$(g) → $H_{Pt}$-$CO_{2(Pt-int)}$-$OH_{int}$+ CO(g) + $H_2O$(g) | 1.138 | 1.692 | 2.124 | 1.564 |
| **S18** | $H_{Pt}$-$CO_{2(Pt-int)}$-$OH_{int}$+ CO(g) + $H_2O$(g) | 0.443 | 0.941 | 0.685 | 1.163 |
| **TS19** | $H_{Pt}$-$CO_{2(Pt-int)}$-$OH_{int}$+ $*_{Pt}$ + CO(g) + $H_2O$(g) →2$H_{Pt}$-$CO_{2(Pt-int)}$+$O_{int}$+ CO(g) + $H_2O$(g) | 0.721 | 1.154 | 1.171 | 2.022 |
| **S19** | 2$H_{Pt}$-$CO_{2(Pt-int)}$+$O_{int}$+ CO(g) + $H_2O$(g) | 0.318 | 0.884 | 0.638 | 1.011 |
| **TS20** | 2$H_{Pt}$-$CO_{2(Pt-int)}$ + CO(g) + $H_2O$(g) → $CO_{2(Pt-int)}$ + 2$*_{Pt}$ + $H_2$(g) + CO(g) + $H_2O$(g) | 0.318 | 0.884 | 0.638 | 1.011 |
| **S03** | $CO_{2(Pt-int)}$ + 2$*_{Pt}$ + $H_2$(g) + CO(g) + $H_2O$(g) | -0.063 | 0.271 | 0.279 | 0.155 |
| **TS21** | $CO_{Pt}$-2$OH_{int}$ + CO(g) + $H_2O$(g) → $COOH_{(Pt\text{-}int)}$-$OH_{int}$+ CO(g) + $H_2O$(g) | 0.997 | 1.513 | 1.994 | 1.320 |
| **S20** | $COOH_{(Pt\text{-}int)}$-$OH_{int}$+ CO(g) + $H_2O$(g) | 1.001 | 1.533 | 1.938 | 1.355 |
| **TS22** | $COOH_{(Pt\text{-}int)}$-$OH_{int}$+ CO(g) + $H_2O$(g) → $CO_{2(Pt\text{-}int)}$-$OH_s$-$OH_{int}$+ CO(g) + $H_2O$(g) | 1.118 | 1.736 | 2.176 | 1.841 |
| **S21** | $CO_{2(Pt\text{-}int)}$-$OH_s$-$OH_{int}$+ CO(g) + $H_2O$(g) | 0.958 | 1.561 | 1.917 | 1.318 |
| **TS23** | $CO_{2(Pt\text{-}int)}$-$OH_s$-$OH_{int}$+ $O_{int}$ + CO(g) + $H_2O$(g) → $CO_{2(Pt\text{-}int)}$-$OH_{int}$-$OH_{int}$+ $O_s$ + CO(g) + $H_2O$(g) | 1.345 | 1.953 | 2.420 | 1.962 |
| **S22** | $CO_{2(Pt\text{-}int)}$-$OH_{int}$-$OH_{int}$+ $O_s$ + CO(g) + $H_2O$(g) | 0.552 | 1.151 | 1.381 | 0.672 |
| **TS24** | $CO_{2(Pt\text{-}int)}$-$OH_{int}$-$OH_{int}$+ $*_{Pt}$ + CO(g) + $H_2O$(g) → $H_{Pt}$-$CO_{2(Pt\text{-}int)}$-$OH_{int}$+$O_{int}$+ CO(g) + $H_2O$(g) | 0.563 | 1.174 | 1.565 | 1.646 |
| **S18** | $H_{Pt}$-$CO_{2(Pt\text{-}int)}$-$OH_{int}$+$O_{int}$+ CO(g) + $H_2O$(g) | 0.444 | 0.941 | 0.685 | 1.163 |

(c) 573 K

| **Label** | **Intermediate or Transition State** | **G (eV)** | | | |
|---|---|---|---|---|---|
| | | **PBE** | **RPBE** | **HSE** | **M06L** |
| **S01** | $*_{Pt}$-$O_{int}$ + 2CO(g) + $H_2O$(g) | 0.000 | 0.000 | 0.000 | 0.000 |
| **TS01** | $*_{Pt}$-$O_{int}$ + 2CO(g) + $H_2O$(g)→ $CO_{Pt}$-$O_{int}$ + CO(g) + $H_2O$(g) | 0.000 | 0.000 | 0.000 | 0.000 |
| **S02** | $CO_{Pt}$-$O_{int}$ + CO(g) + $H_2O$(g) | 0.052 | 0.143 | 0.235 | 0.638 |
| **TS02** | $CO_{Pt}$-$O_{int}$ + CO(g) + $H_2O$(g) → $CO_{2(Pt\text{-}int)}$ + CO(g) + $H_2O$(g) | 0.704 | 0.922 | 0.911 | 1.37 |
| **S03** | $CO_{2(Pt\text{-}int)}$ + CO(g) + $H_2O$(g) | 0.549 | 0.836 | 0.692 | 0.901 |
| **TS03** | $CO_{2(Pt\text{-}int)}$ + CO(g) + $H_2O$(g) → $*_{Pt}$-$V_{int}$ + $CO_2$(g) + CO(g) + $H_2O$(g) | 1.305 | 1.452 | 2.279 | 1.571 |
| **S04** | $*_{Pt}$-$V_{int}$ + CO(g) + $H_2O$(g) + $CO_2$(g) | 0.547 | 0.466 | 0.926 | 1.132 |
| **TS04** | $*_{Pt}$-$V_{int}$ + CO(g) + $H_2O$(g) + $CO_2$(g)→ $*_{Pt}$-2$OH_{int}$ + CO(g) + $CO_2$(g) | 1.086 | 1.329 | 2.110 | 1.461 |
| **S06** | $*_{Pt}$-2$OH_{int}$ + CO(g) + $CO_2$(g) | 0.144 | 0.485 | 0.690 | -0.330 |
| **TS05** | $*_{Pt}$-2$OH_{int}$ + CO(g) + $CO_2$(g) → $H_{Pt}$-$OH_{int}$ + $O_{int}$ + CO(g) + $CO_2$(g) | 0.753 | 0.961 | 1.203 | 1.405 |
| **S07** | $H_{Pt}$-$OH_{int}$ + $*_{Pt}$ + CO(g) + $CO_2$(g) | 0.409 | 0.542 | 0.771 | 0.877 |
| **TS06** | $H_{Pt}$-$OH_{int}$ + $*_{Pt}$ + CO(g) + $CO_2$(g) → $2H_{Pt}$-$O_{int}$ + CO(g) + $CO_2$(g) | 0.989 | 1.377 | 1.806 | 1.292 |
| **S08** | $2H_{Pt}$-$O_{int}$ + CO(g) + $CO_2$(g) | -0.062 | 0.281 | 0.345 | 0.211 |
| **TS07** | $2H_{Pt}$-$O_{int}$ + CO(g) + $CO_2$(g) → $*_{Pt}$-$O_{int}$ +$*_{Pt}$ +$H_2$(g) + CO(g) + $CO_2$(g) | -0.062 | 0.281 | 0.345 | 0.211 |
| **S01** | $*_{Pt}$-$O_{int}$ +$*_{Pt}$ +$H_2$(g) + CO(g) + $CO_2$(g) | -0.651 | -0.606 | -0.452 | -0.787 |
| **TS08** | $CO_{2(Pt-int)}$ + CO(g) + $H_2O$(g) → $CO_{Pt}$-$CO_{2(int)}$ + $H_2O$(g) | 1.050 | 1.337 | 1.193 | 1.402 |
| **S09** | $CO_{Pt}$-$CO_{2(int)}$ + $H_2O$(g) | 0.271 | 0.511 | 0.610 | 1.049 |
| **TS09** | $CO_{Pt}$-$CO_{2(int)}$ + $H_2O$(g) → $CO_{Pt}$-$V_{int}$ + $CO_2$(g) + $H_2O$(g) | 0.640 | 0.691 | 1.219 | 1.674 |
| **S10** | $CO_{Pt}$-$V_{int}$ +$O_{int}$ + $H_2O$(g) + $CO_2$(g) | -0.355 | -0.313 | 0.557 | 0.447 |
| **TS10** | $CO_{Pt}$-$V_{int}$ +$O_{int}$ $H_2O$(g) + $CO_2$(g) → $CO_{Pt}$-$2OH_{int}$ + $CO_2$(g) | 0.839 | 0.965 | 1.411 | 1.873 |
| **S11** | $CO_{Pt}$-$2OH_{int}$ + $O_s$ + $CO_2$(g) | 0.193 | 0.553 | 0.837 | 0.450 |
| **TS11** | $CO_{Pt}$-$2OH_{int}$ + $O_s$ + $CO_2$(g) → $CO_{Pt}$-$OH_{int}$-$O_{int}$-$OH_s$ + $CO_2$(g) | 0.691 | 1.031 | 1.126 | 1.571 |
| **S12** | $CO_{Pt}$-$OH_{int}$-$O_{int}$-$OH_s$ + $O_{int}$ + $CO_2$ (g) | 0.285 | 0.477 | 0.564 | 1.000 |
| **TS12** | $CO_{Pt}$-$OH_{int}$-$O_{int}$-$OH_s$ + $O_{int}$ + $CO_2$ (g) → $CO_{Pt}$-$OH_{int}$-$O_{int}$-$OH_{int}$ + $O_s$ + $CO_2$ (g) | 0.830 | 0.910 | 1.219 | 1.881 |
| **S13** | $CO_{Pt}$-$OH_{int}$-$O_{int}$-$OH_{int}$ + $*_{Pt}$ + $CO_2$ (g) | 0.171 | 0.452 | 0.712 | 0.449 |
| **TS13** | $CO_{Pt}$-$OH_{int}$-$O_{int}$-$OH_{int}$ + $*_{Pt}$ + $CO_2$ (g)→ $CO_{Pt}$-$OH_{int}$-$O_{int}$-$H_{Pt}$ + $O_{int}$ + $CO_2$ (g) | 0.199 | 0.334 | 0.585 | 1.281 |
| **S14** | $CO_{Pt}$-$OH_{int}$-$O_{int}$-$H_{Pt}$ + $*_{Pt}$ + $CO_2$ (g) | -0.412 | -0.145 | -0.222 | 0.171 |
| **TS14** | $CO_{Pt}$-$OH_{int}$-$O_{int}$-$H_{Pt}$ + $*_{Pt}$ + $CO_2$ (g)→ $CO_{Pt}$-$O_{int}$-$2H_{Pt}$ + $O_{int}$ + $CO_2$ (g) | -0.120 | 0.063 | 0.211 | 1.021 |
| **S15** | $CO_{Pt}$-$O_{int}$-$2H_{Pt}$ + $CO_2$ (g) | -0.799 | -0.521 | -0.650 | 0.064 |
| **TS15** | $CO_{Pt}$-$O_{int}$-$2H_{Pt}$ + $CO_2$ (g)→ $CO_{Pt}$-$O_{int}$ + $2*_{Pt}$ + $CO_2$ (g)+ $H_2$ (g) | -0.799 | -0.521 | -0.650 | 0.064 |
| **S02** | $CO_{Pt}$-$O_{int}$ + $2*_{Pt}$ + $CO_2$ (g)+ $H_2$ (g) | -0.599 | -0.463 | -0.217 | -0.148 |
| **TS16** | $CO_{Pt}$-$2OH_{int}$ + CO(g) + $H_2O$(g) →$CHO_{Pt}$-$O_{int}$-$OH_{int}$+ CO(g) + $H_2O$(g) | 1.006 | 1.379 | 1.853 | 2.002 |
| **S16** | $CHO_{Pt}$-$O_{int}$-$OH_{int}$+ CO(g) + $H_2O$(g) | 0.443 | 0.787 | 0.994 | 1.234 |
| **TS17** | $CHO_{Pt}$-$O_{int}$-$OH_{int}$+ CO(g) + $H_2O$(g) → $*_{Pt}$ –$HCOO_{int}$-$OH_{int}$+ CO(g) + $H_2O$(g) | 1.069 | 1.666 | 1.751 | 1.772 |
| **S17** | $*_{Pt}$ –$HCOO_{int}$-$OH_{int}$+ CO(g) + $H_2O$(g) | 0.505 | 1.149 | 1.466 | 0.576 |
| **TS18** | $*_{Pt}$ –$HCOO_{int}$-$OH_{int}$ + $*_{Pt}$ + CO(g) + $H_2O$(g) → $H_{Pt}$-$CO_{2(Pt\text{-}int)}$-$OH_{int}$+ CO(g) + $H_2O$(g) | 1.144 | 1.698 | 2.130 | 1.570 |
| **S18** | $H_{Pt}$-$CO_{2(Pt\text{-}int)}$-$OH_{int}$+ CO(g) + $H_2O$(g) | 0.426 | 0.924 | 0.668 | 1.146 |
| **TS19** | $H_{Pt}$-$CO_{2(Pt\text{-}int)}$-$OH_{int}$+ $*_{Pt}$ + CO(g) + $H_2O$(g) →2$H_{Pt}$-$CO_{2(Pt\text{-}int)}$+$O_{int}$+ CO(g) + $H_2O$(g) | 0.712 | 1.145 | 1.162 | 2.013 |
| **S19** | 2$H_{Pt}$-$CO_{2(Pt\text{-}int)}$+$O_{int}$+ CO(g) + $H_2O$(g) | 0.298 | 0.864 | 0.618 | 0.991 |
| **TS20** | 2$H_{Pt}$-$CO_{2(Pt\text{-}int)}$ + CO(g) + $H_2O$(g) → $CO_{2(Pt\text{-}int)}$ + 2$*_{Pt}$ + $H_2$(g) + CO(g) + $H_2O$(g) | 0.298 | 0.864 | 0.618 | 0.991 |
| **S03** | $CO_{2(Pt\text{-}int)}$ + 2$*_{Pt}$ + $H_2$(g) + CO(g) + $H_2O$(g) | -0.102 | 0.233 | 0.241 | 0.117 |
| **TS21** | $CO_{Pt}$-2$OH_{int}$ + CO(g) + $H_2O$(g) → $COOH_{(Pt\text{-}int)}$-$OH_{int}$+ CO(g) + $H_2O$(g) | 0.967 | 1.483 | 1.964 | 1.290 |
| **S20** | $COOH_{(Pt\text{-}int)}$-$OH_{int}$+ CO(g) + $H_2O$(g) | 0.917 | 1.449 | 1.854 | 1.271 |
| **TS22** | $COOH_{(Pt\text{-}int)}$-$OH_{int}$+ CO(g) + $H_2O$(g) → $CO_{2(Pt\text{-}int)}$-$OH_s$-$OH_{int}$+ CO(g) + $H_2O$(g) | 1.086 | 1.704 | 2.144 | 1.809 |
| **S21** | $CO_{2(Pt\text{-}int)}$-$OH_s$-$OH_{int}$+ CO(g) + $H_2O$(g) | 0.910 | 1.513 | 1.869 | 1.270 |
| **TS23** | $CO_{2(Pt\text{-}int)}$-$OH_s$-$OH_{int}$+ $O_{int}$ + CO(g) + $H_2O$(g) → $CO_{2(Pt\text{-}int)}$-$OH_{int}$-$OH_{int}$+ $O_s$ + CO(g) + $H_2O$(g) | 1.312 | 1.920 | 2.387 | 1.929 |
| **S22** | $CO_{2(Pt\text{-}int)}$-$OH_{int}$-$OH_{int}$+ $O_s$ + CO(g) + $H_2O$(g) | 0.501 | 1.100 | 1.330 | 0.621 |
| **TS24** | $CO_{2(Pt\text{-}int)}$-$OH_{int}$-$OH_{int}$+ $*_{Pt}$ + CO(g) + $H_2O$(g) → $H_{Pt}$-$CO_{2(Pt\text{-}int)}$-$OH_{int}$+$O_{int}$+ CO(g) + $H_2O$(g) | 0.516 | 1.127 | 1.518 | 1.599 |
| **S18** | $H_{Pt}$-$CO_{2(Pt\text{-}int)}$-$OH_{int}$+$O_{int}$+ CO(g) + $H_2O$(g) | 0.426 | 0.923 | 0.667 | 1.145 |

Supplementary Table 2. Relative free energies as calculated by four DFT functionals[7-12] for Pt(111).

(a) 503 K

| | ΔG/ ΔG‡ (eV) | | | |
|---|---|---|---|---|
| **Reaction** | **PBE** | **RPBE** | **M06L** | **HSE** |
| Vacant site - clean Pt(111) | 0.000 | 0.000 | 0.000 | 0.000 |
| $CO(g)+^{*}\leftrightarrow CO^{*}\ \Delta G$ | -1.000 | -0.601 | -0.666 | -0.991 |
| $H_2O(g)+^{*}\leftrightarrow H_2O^{*}\ \Delta G$ | 0.524 | 0.748 | 0.507 | 0.410 |
| $H_2O^{*}\leftrightarrow H^{*}+OH^{*}\ \Delta G$ | 0.473 | 0.996 | 1.472 | 0.618 |
| $H_2O^{*}\leftrightarrow H^{*}+OH^{*}\ \Delta G^{\ddagger}$ | 0.794 | 1.390 | 1.072 | 0.897 |
| $OH^{*}\leftrightarrow\ H^{*}+O^{*}\ \Delta G$ | -0.117 | 0.297 | 1.152 | 0.429 |
| $OH^{*}\leftrightarrow\ H^{*}+O^{*}\ \Delta G^{\ddagger}$ | 0.870 | 0.905 | 1.184 | 1.290 |
| $OH^{*}+OH^{*}\leftrightarrow H_2O^{*}+O^{*}\ \ \Delta G$ | -0.584 | -0.694 | -0.315 | -0.184 |
| $OH^{*}+OH^{*}\leftrightarrow H_2O^{*}+O^{*}\ \ \Delta G^{\ddagger}$ | 0.000 | 0.000 | 0.000 | 0.000 |
| $CO^{*}+\ O^{*}\leftrightarrow CO_2^{*}\ +*\ \ \Delta G$ | -0.274 | -1.150 | -1.608 | -1.329 |
| $CO^{*}+\ O^{*}\leftrightarrow CO_2^{*}\ +*\ \ \Delta G^{\ddagger}$ | 1.042 | 0.922 | 0.894 | 1.267 |
| $CO^{*}+OH^{*}\leftrightarrow\ COOH^{*}+*\ \ \Delta G$ | -0.110 | -0.577 | -0.806 | -0.391 |
| $CO^{*}+OH^{*}\leftrightarrow\ COOH^{*}+*\ \ \Delta G^{\ddagger}$ | 0.414 | 0.303 | 0.312 | 2.739 |
| $COOH^{*}+*\leftrightarrow CO_2^{*}+\ H^{*}\ \Delta G$ | -0.324 | -0.319 | 0.307 | -0.552 |
| $COOH^{*}+*\leftrightarrow CO_2^{*}+\ H^{*}\ \Delta G^{\ddagger}$ | 0.613 | -0.277 | 0.732 | -0.001 |
| $COOH^{*}+O^{*}\leftrightarrow CO_2^{*}+OH^{*}\ \Delta G$ | 0.077 | -0.331 | -0.561 | -0.697 |
| $COOH^{*}+O^{*}\leftrightarrow CO_2^{*}+OH^{*}\ \Delta G^{\ddagger}$ | 0.538 | 0.587 | 0.546 | 0.536 |
| $COOH^{*}+OH^{*}\leftrightarrow CO_2^{*}+H_2O^{*}\ \Delta G$ | -0.695 | -1.213 | -1.063 | -1.069 |
| $COOH^{*}+OH^{*}\leftrightarrow CO_2^{*}+H_2O^{*}\ \Delta G^{\ddagger}$ | 0.000 | 0.000 | 1.857 | 0.000 |
| $HCOO^{*}+*\leftrightarrow CO_2^{*}+H^{*}\ \Delta G$ | -0.628 | -0.741 | 0.099 | -0.484 |
| $HCOO^{*}+*\leftrightarrow CO_2^{*}+H^{*}\ \Delta G^{\ddagger}$ | 0.970 | 1.000 | 0.987 | 1.038 |
| $HCOO^{*}+O^{*}\leftrightarrow CO_2^{*}+OH^{*}\ \Delta G$ | -0.514 | -1.041 | -1.058 | -0.917 |
| $HCOO^{*}+O^{*}\leftrightarrow CO_2^{*}+OH^{*}\ \Delta G^{\ddagger}$ | 1.220 | 1.118 | 1.392 | 1.727 |
| $HCOO^* + OH^* \leftrightarrow CO_2^* + H_2O^*$ $\Delta G$ | -1.004 | -1.640 | -1.277 | -1.005 |
| $HCOO^* + OH^* \leftrightarrow CO_2^* + H_2O^*$ $\Delta G^\ddagger$ | 0.899 | 0.908 | 1.122 | 1.303 |
| $2H^* \leftrightarrow H_2 + 2*$ | 0.398 | 0.733 | 1.035 | 0.161 |
| $CO_2^* \leftrightarrow CO_2 + *$ | -0.783 | -0.769 | -0.805 | -0.777 |
| $HCO^* + * \leftrightarrow CO^* + H^*$ $\Delta G$ | -1.138 | -0.678 | -0.207 | -1.426 |
| $HCO^* + * \leftrightarrow CO^* + H^*$ $\Delta G^\ddagger$ | 0.086 | 0.158 | 0.246 | 0.125 |

(b) 523 K

| | **ΔG/ ΔG‡ (eV)** | | | |
|---|---|---|---|---|
| **Reaction** | **PBE** | **RPBE** | **M06L** | **HSE** |
| Vacant site - clean Pt(111) | 0.000 | 0.000 | 0.000 | 0.000 |
| $CO(g) + {}^* \leftrightarrow CO^*$ $\Delta G$ | -0.970 | -0.571 | -0.636 | -0.961 |
| $H_2O(g) + {}^* \leftrightarrow H_2O^*$ $\Delta G$ | 0.550 | 0.774 | 0.533 | 0.436 |
| $H_2O^* \leftrightarrow H^* + OH^*$ $\Delta G$ | 0.475 | 0.998 | 1.474 | 0.621 |
| $H_2O^* \leftrightarrow H^* + OH^*$ $\Delta G^\ddagger$ | 0.794 | 1.390 | 1.072 | 0.897 |
| $OH^* \leftrightarrow H^* + O^*$ $\Delta G$ | -0.116 | 0.298 | 1.153 | 0.430 |
| $OH^* \leftrightarrow H^* + O^*$ $\Delta G^\ddagger$ | 0.869 | 0.904 | 1.183 | 1.289 |
| $OH^* + OH^* \leftrightarrow H_2O^* + O^*$ $\Delta G$ | -0.585 | -0.694 | -0.315 | -0.184 |
| $OH^* + OH^* \leftrightarrow H_2O^* + O^*$ $\Delta G^\ddagger$ | -0.038 | 0.000 | 0.000 | 0.000 |
| $CO^* + O^* \leftrightarrow CO_2^* + *$ $\Delta G$ | -0.278 | -1.154 | -1.612 | -1.333 |
| $CO^* + O^* \leftrightarrow CO_2^* + *$ $\Delta G^\ddagger$ | 1.041 | 0.921 | 0.893 | 1.266 |
| $CO^* + OH^* \leftrightarrow COOH^* + *$ $\Delta G$ | -0.110 | -0.577 | -0.806 | -0.391 |
| $CO^* + OH^* \leftrightarrow COOH^* + *$ $\Delta G^\ddagger$ | 0.414 | 0.303 | 0.312 | 2.739 |
| $COOH^* + * \leftrightarrow CO_2^* + H^*$ $\Delta G$ | -0.328 | -0.323 | 0.303 | -0.556 |
| $COOH^* + * \leftrightarrow CO_2^* + H^*$ $\Delta G^\ddagger$ | 0.614 | -0.277 | 0.733 | 0.000 |
| $COOH^* + O^* \leftrightarrow CO_2^* + OH^*$ $\Delta G$ | 0.083 | -0.325 | -0.555 | -0.691 |
| $COOH^* + O^* \leftrightarrow CO_2^* + OH^*$ $\Delta G^\ddagger$ | 0.550 | 0.599 | 0.558 | 0.548 |
| $COOH^* + OH^* \leftrightarrow CO_2^* + H_2O^*$ $\Delta G$ | -0.696 | -1.214 | -1.064 | -1.070 |
| $COOH^* + OH^* \leftrightarrow CO_2^* + H_2O^*\ \Delta G^‡$ | 0.000 | 0.000 | 1.857 | 0.000 |
| $HCOO^* + * \leftrightarrow CO_2^* + H^*\ \Delta G$ | -0.630 | -0.743 | 0.096 | -0.486 |
| $HCOO^* + * \leftrightarrow CO_2^* + H^*\ \Delta G^‡$ | 0.969 | 0.999 | 0.986 | 1.037 |
| $HCOO^* + O^* \leftrightarrow CO_2^* + OH^*\ \Delta G$ | -0.517 | -1.044 | -1.060 | -0.919 |
| $HCOO^* + O^* \leftrightarrow CO_2^* + OH^*\ \Delta G^‡$ | 1.218 | 1.116 | 1.390 | 1.725 |
| $HCOO^* + OH^* \leftrightarrow CO_2^* + H_2O^*\ \Delta G$ | -1.003 | -1.640 | -1.277 | -1.005 |
| $HCOO^* + OH^* \leftrightarrow CO_2^* + H_2O^*\ \Delta G^‡$ | 0.898 | 0.907 | 1.121 | 1.302 |
| $2H^* \leftrightarrow H_2 + 2*$ | 0.376 | 0.711 | 1.013 | 0.139 |
| $CO_2^* \leftrightarrow CO_2 + *$ | -0.812 | -0.798 | -0.834 | -0.806 |
| $HCO^* + * \leftrightarrow CO^* + H^*\ \Delta G$ | -1.137 | -0.677 | -0.206 | -1.425 |
| $HCO^* + * \leftrightarrow CO^* + H^*\ \Delta G^‡$ | 0.085 | 0.157 | 0.245 | 0.124 |

(b) 573 K

| | ΔG/ ΔG‡ (eV) | | | |
|---|---|---|---|---|
| **Reaction** | **PBE** | **RPBE** | **M06L** | **HSE** |
| Vacant site - clean Pt(111) | 0.000 | 0.000 | 0.000 | 0.000 |
| $CO(g) + ^* \leftrightarrow CO^*\ \Delta G$ | -0.895 | -0.496 | -0.561 | -0.886 |
| $H_2O(g) + ^* \leftrightarrow H_2O^*\ \Delta G$ | 0.615 | 0.839 | 0.598 | 0.501 |
| $H_2O^* \leftrightarrow H^* + OH^*\ \Delta G$ | 0.481 | 1.004 | 1.480 | 0.626 |
| $H_2O^* \leftrightarrow H^* + OH^*\ \Delta G^‡$ | 0.797 | 1.393 | 1.075 | 0.900 |
| $OH^* \leftrightarrow H^* + O^*\ \Delta G$ | -0.113 | 0.301 | 1.156 | 0.433 |
| $OH^* \leftrightarrow H^* + O^*\ \Delta G^‡$ | 0.867 | 0.902 | 1.181 | 1.287 |
| $OH^* + OH^* \leftrightarrow H_2O^* + O^*\ \Delta G$ | -0.586 | -0.696 | -0.317 | -0.186 |
| $OH^* + OH^* \leftrightarrow H_2O^* + O^*\ \Delta G^‡$ | -0.037 | 0.000 | 0.000 | 0.000 |
| $CO^* + O^* \leftrightarrow CO_2^* + *\ \Delta G$ | -0.287 | -1.163 | -1.621 | -1.342 |
| $CO^* + O^* \leftrightarrow CO_2^* + *\ \Delta G^‡$ | 1.039 | 0.919 | 0.891 | 1.264 |
| $CO^* + OH^* \leftrightarrow COOH^* + *\ \Delta G$ | -0.110 | -0.577 | -0.806 | -0.391 |
| $CO^* + OH^* \leftrightarrow COOH^* + *\ \Delta G^‡$ | 0.412 | 0.301 | 0.310 | 2.737 |
| $COOH^* + * \leftrightarrow CO_2^* + H^*\ \Delta G$ | -0.245 | -0.240 | 0.386 | -0.473 |
| $COOH^* + * \leftrightarrow CO_2^* + H^*\ \Delta G^\ddagger$ | 0.496 | -0.394 | 0.615 | -0.118 |
| $COOH^* + O^* \leftrightarrow CO_2^* + OH^*\ \Delta G$ | -0.116 | -0.525 | -0.755 | -0.891 |
| $COOH^* + O^* \leftrightarrow CO_2^* + OH^*\ \Delta G^\ddagger$ | 0.610 | 0.659 | 0.618 | 0.608 |
| $COOH^* + OH^* \leftrightarrow CO_2^* + H_2O^*\ \Delta G$ | -0.687 | -1.205 | -1.056 | -1.061 |
| $COOH^* + OH^* \leftrightarrow CO_2^* + H_2O^*\ \Delta G^\ddagger$ | 0.000 | 0.000 | 1.857 | 0.000 |
| $HCOO^* + * \leftrightarrow CO_2^* + H^*\ \Delta G$ | -0.673 | -0.786 | 0.053 | -0.529 |
| $HCOO^* + * \leftrightarrow CO_2^* + H^*\ \Delta G^\ddagger$ | 0.966 | 0.996 | 0.983 | 1.034 |
| $HCOO^* + O^* \leftrightarrow CO_2^* + OH^*\ \Delta G$ | -0.451 | -0.978 | -0.994 | -0.853 |
| $HCOO^* + O^* \leftrightarrow CO_2^* + OH^*\ \Delta G^\ddagger$ | 1.211 | 1.109 | 1.383 | 1.718 |
| $HCOO^* + OH^* \leftrightarrow CO_2^* + H_2O^*\ \Delta G$ | -1.003 | -1.639 | -1.277 | -1.005 |
| $HCOO^* + OH^* \leftrightarrow CO_2^* + H_2O^*\ \Delta G^\ddagger$ | 0.893 | 0.902 | 1.116 | 1.297 |
| $2H^* \leftrightarrow H_2 + 2 *$ | 0.308 | 0.643 | 0.945 | 0.071 |
| $CO_2^* \leftrightarrow CO_2 + *$ | -0.880 | -0.866 | -0.902 | -0.874 |
| $HCO^* + * \leftrightarrow CO^* + H^*\ \Delta G$ | -1.136 | -0.676 | -0.205 | -1.424 |
| $HCO^* + * \leftrightarrow CO^* + H^*\ \Delta G^\ddagger$ | 0.084 | 0.156 | 0.244 | 0.123 |

Supplementary Table 3. Relative free energies as calculated by four DFT functionals[7-12] for the interface corner active site.[13]

(a) 503 K

| Intermediate or Transition State | ΔG (eV) | | | |
|---|---|---|---|---|
| | PBE | RPBE | HSE | M06L |
| $*_{Pt}$ (IM1) + $2CO_{(g)}$ + $H_2O_{(g)}$ | 0.000 | 0.000 | 0.000 | 0.000 |
| $*_{Pt}$ (IM1) + $2CO_{(g)}$ + $H_2O_{(g)} \rightarrow CO_{Pt}$ (IM2) + $CO_{(g)}$ + $H_2O_{(g)}$ | 0.000 | 0.000 | 0.000 | 0.000 |
| $CO_{Pt}$ (IM2) + $*_{Ti}$ + $CO_{(g)}$ + $H_2O_{(g)}$ | 0.046 | 0.320 | 0.045 | 0.500 |
| $CO_{Pt}$ (IM2) + $*_{Ti}$ + $CO_{(g)}$ + $H_2O_{(g)} \rightarrow CO_{Pt}$-$H_2O_{Ti}$ (IM3) + $CO_{(g)}$ | 0.046 | 0.320 | 0.045 | 0.500 |
| $CO_{Pt}$-$H_2O_{Ti}$ (IM3) + $O_b$ + $CO_{(g)}$ | 0.482 | 1.014 | 0.482 | 0.647 |
| $CO_{Pt}$-$H_2O_{Ti}$ (IM3) + $O_b$ + $CO_{(g)} \rightarrow CO_{Pt}$-$OH_{Ti}$-$O_bH$ (IM4) + $CO_{(g)}$ | 1.140 | 1.806 | 1.161 | 1.459 |
| $CO_{Pt}$-$OH_{Ti}$-$O_bH$ (IM4) + $CO_{(g)}$ | 0.480 | 1.070 | 0.402 | 0.475 |
| $CO_{Pt}$-$OH_{Ti}$-$O_bH$ (IM4) + $CO_{(g)} \rightarrow COOH_{(Pt-Ti)}$-$O_bH$ (IM5) + $CO_{(g)}$ | 0.947 | 1.718 | 0.939 | 1.087 |
| $COOH_{(Pt-Ti)}$-$O_bH$ (IM5)+ $O_s$+ $CO_{(g)}$ | 0.127 | 0.915 | 0.066 | 0.375 |
| $COOH_{(Pt-Ti)}$-$O_bH$ (IM5)+ $O_s$+ $CO_{(g)} \rightarrow CO_{2(Pt-Ti)}$-$O_bH$-$O_sH$ (IM6) + $CO_{(g)}$ | 0.656 | 1.545 | 0.748 | 1.401 |
| $CO_{2(Pt-Ti)}$-$O_bH$-$O_sH$ (IM6) + $CO_{(g)}$ | 0.359 | 1.267 | 0.378 | 0.678 |
| $CO_{2(Pt-Ti)}$-$O_bH$-$O_sH$ (IM6) + $CO_{(g)} \rightarrow *_{Pt}$-$O_bH$-$O_sH$ (IM7) + $*_{Ti}$ + $CO_{(g)}$ + $CO_{2(g)}$ | 0.359 | 1.267 | 0.378 | 0.678 |
| $*_{Pt}$-$O_bH$-$O_sH$ (IM7) + $*_{Pt}$ + $CO_{(g)}$ + $CO_{2(g)}$ | -0.054 | 0.404 | 0.495 | -0.078 |
| $*_{Pt}$-$O_bH$-$O_sH$ (IM7) + $*_{Pt}$ + $CO_{(g)}$ + $CO_{2(g)} \rightarrow$ $*_{Pt}$-$H_{Pt}$-$O_bH$ (IM8) + $O_s$+ $CO_{(g)}$ + $CO_{2(g)}$ | -0.195 | 0.247 | 0.652 | 0.292 |
| $*_{Pt}$-$H_{Pt}$-$O_bH$ (IM8) + $O_s$+ $CO_{(g)}$ + $CO_{2(g)}$ | -0.571 | -0.237 | -0.425 | -0.342 |
| $*_{Pt}$-$H_{Pt}$-$O_bH$ (IM8) + $O_s$ + $CO_{(g)}$ + $CO_{2(g)} \rightarrow *_{Pt}$-$H_{Pt}$-$O_sH$ (IM9) + $O_b$+ $CO_{(g)}$ + $CO_{2(g)}$ | -0.138 | 0.247 | 0.116 | 0.771 |
| $*_{Pt}$-$H_{Pt}$-$O_sH$ (IM9) + $CO_{(g)}$ + $CO_{2(g)}$ | -0.816 | -0.469 | -0.630 | -0.229 |
| $*_{Pt}$-$H_{Pt}$-$O_sH$ (IM9) + $CO_{(g)}$ + $CO_{2(g)} \rightarrow 2H_{Pt}$ (IM10) + $O_s$+ $CO_{(g)}$ + $CO_{2(g)}$ | -0.497 | -0.173 | -0.172 | 0.397 |
| $2H_{Pt}$ (IM10) + $CO_{(g)}$ + $CO_{2(g)}$ | -1.365 | -1.102 | -1.303 | -0.748 |
| $2H_{Pt}$ (IM10) + $CO_{(g)}$ + $CO_{2(g)} \rightarrow *_{Pt}$ (IM1) + $*_{Pt}$+ $CO_{(g)}$ + $CO2_{(g)}$ + $H_{2(g)}$ | -1.365 | -1.102 | -1.303 | -0.748 |
| $*_{Pt}$ (IM1) + $*_{Pt}$+ $CO_{(g)}$ + $CO_{2(g)}$ + $H_{2(g)}$ | -0.884 | -0.836 | -0.686 | -1.018 |
| $CO_{Pt}$ (IM2) + $CO_{(g)}$ + $H_2O_{(g)}$→ $(CO,CO)_{Pt}$ (IM11) + $H_2O_{(g)}$ | 0.477 | 0.751 | 0.476 | 0.931 |
| $(CO,CO)_{Pt}$ (IM11) + $O_i$ + $H_2O_{(g)}$ | 0.126 | 0.185 | 0.261 | -0.171 |
| $(CO,CO)_{Pt}$ (IM11) + $O_i$ + $H_2O_{(g)}$→ $(CO,CO_2)_{Pt-Oi}$ (IM12) + $H_2O_{(g)}$ | 0.599 | 0.869 | 0.793 | 0.359 |
| $(CO,CO_2)_{Pt-Oi}$ (IM12) + $H_2O_{(g)}$ | 0.026 | 0.306 | 0.253 | -0.227 |
| $(CO,CO_2)_{Pt-Oi}$ (IM12) + $H_2O_{(g)}$→ $CO_{Pt}$-$V_i$ (IM13) + $H_2O_{(g)}$+ $CO_{2(g)}$ | 0.046 | 0.184 | 1.229 | -0.599 |
| $CO_{Pt}$-$V_i$ (IM13) + $H_2O_{(g)}$+ $CO_{2(g)}$ | -0.485 | -0.652 | 0.805 | -1.045 |
| $CO_{Pt}$-$V_i$ (IM13) + $H_2O_{(g)}$+ $CO_{2(g)}$→ $CO_{Pt}$-$H_2O_i$ (IM14) + $CO_{2(g)}$ | -0.485 | -0.652 | 0.805 | -1.045 |
| $CO_{Pt}$-$H_2O_i$ (IM14) + $*_{Pt}$ + $CO_{2(g)}$ | -0.825 | -0.652 | 0.304 | -1.155 |
| $CO_{Pt}$-$H_2O_i$ (IM14) + $*_{Pt}$ + $CO_{2(g)}$ → $CO_{Pt}$-$H_{Pt}$-$O_iH$ (IM15) + $CO_{2(g)}$ | -0.303 | -0.158 | 0.468 | -0.600 |
| $CO_{Pt}$-$H_{Pt}$-$O_iH$ (IM15) + $CO_{2(g)}$ | -0.649 | -0.431 | -0.421 | -0.873 |
| $CO_{Pt}$-$H_{Pt}$-$O_iH$ (IM15) + $CO_{2(g)}$→ $(CO,H)_{Pt}$-$H_{Pt}$ (IM16) + $O_i$ + $CO_{2(g)}$ | 0.059 | 0.268 | 0.336 | 0.014 |
| $(CO,H)_{Pt}$-$H_{Pt}$ (IM16) + $CO_{2(g)}$ | -0.430 | -0.336 | -0.304 | -0.629 |
| $(CO,H)_{Pt}$-$H_{Pt}$ (IM16) + $CO_{2(g)}$ → $CO_{Pt}$ (IM2) + $*_{Pt}$+ $CO_{2(g)}$ + $H_2$(g) | -0.430 | -0.336 | -0.304 | -0.629 |
| $CO_{Pt}$ (IM2) + $*_{Pt}$+ $CO_{2(g)}$ + $H_2$(g) | -0.376 | -0.297 | -0.082 | -1.153 |

(b) 523 K

| **Intermediate or Transition State** | **ΔG (eV)** | | | |
|---|---|---|---|---|
| | **PBE** | **RPBE** | **HSE** | **M06L** |
| $*_{Pt}$ (IM1) + $2CO_{(g)}$ + $H_2O_{(g)}$ | 0.000 | 0.000 | 0.000 | 0.000 |
| $*_{Pt}$ (IM1) + $2CO_{(g)}$ + $H_2O_{(g)}$→ $CO_{Pt}$ (IM2) + $CO_{(g)}$ + $H_2O_{(g)}$ | 0.000 | 0.000 | 0.000 | 0.000 |
| $CO_{Pt}$ (IM2) + $*_{Ti}$ + $CO_{(g)}$ + $H_2O_{(g)}$ | -0.468 | -0.194 | -0.469 | -0.014 |
| $CO_{Pt}$ (IM2) + $*_{Ti}$ + $CO_{(g)}$ + $H_2O_{(g)}$→$CO_{Pt}$-$H_2O_{Ti}$ (IM3) + $CO_{(g)}$ | -0.468 | -0.194 | -0.469 | -0.014 |
| $CO_{Pt}$-$H_2O_{Ti}$ (IM3) + $O_b$ + $CO_{(g)}$ | 0.072 | 0.604 | 0.072 | 0.237 |
| $CO_{Pt}$-$H_2O_{Ti}$ (IM3) + $O_b$ + $CO_{(g)}$→$CO_{Pt}$-$OH_{Ti}$-$O_bH$ (IM4) + $CO_{(g)}$ | 0.305 | 0.971 | 0.326 | 0.624 |
| $CO_{Pt}$-$OH_{Ti}$-$O_bH$ (IM4) + $CO_{(g)}$ | 0.266 | 0.856 | 0.188 | 0.261 |
| $CO_{Pt}$-$OH_{Ti}$-$O_bH$ (IM4) + $CO_{(g)}$ → $COOH_{(Pt-Ti)}$-$O_bH$ (IM5) + $CO_{(g)}$ | 0.818 | 1.589 | 0.810 | 0.958 |
| $COOH_{(Pt-Ti)}$-$O_bH$ (IM5)+ $O_s$+ $CO_{(g)}$ | 0.707 | 1.495 | 0.646 | 0.955 |
| $COOH_{(Pt-Ti)}$-$O_bH$ (IM5)+ $O_s$+ $CO_{(g)}$ →$CO_{2(Pt-Ti)}$-$O_bH$-$O_sH$ (IM6) + $CO_{(g)}$ | 1.204 | 2.093 | 1.296 | 1.949 |
| $CO_{2(Pt-Ti)}$-$O_bH$-$O_sH$ (IM6) + $CO_{(g)}$ | 1.178 | 2.086 | 1.196 | 1.497 |
| $CO_{2(Pt-Ti)}$-$O_bH$-$O_sH$ (IM6) + $CO_{(g)}$ → $*_{Pt}$-$O_bH$-$O_sH$ (IM7) + $*_{Ti}$ + $CO_{(g)}$ + $CO_{2(g)}$ | 1.178 | 2.086 | 1.196 | 1.497 |
| $*_{Pt}$-$O_bH$-$O_sH$ (IM7) + $*_{Pt}$ + $CO_{(g)}$ + $CO_{2(g)}$ | 0.529 | 0.987 | 1.078 | 0.505 |
| $*_{Pt}$-$O_bH$-$O_sH$ (IM7) + $*_{Pt}$ + $CO_{(g)}$ + $CO_{2(g)}$ → $*_{Pt}$-$H_{Pt}$-$O_bH$ (IM8) + $O_s$+ $CO_{(g)}$ + $CO_{2(g)}$ | 0.754 | 1.196 | 1.601 | 1.241 |
| $*_{Pt}$-$H_{Pt}$-$O_bH$ (IM8) + $O_s$+ $CO_{(g)}$ + $CO_{2(g)}$ | -0.161 | 0.173 | -0.015 | 0.068 |
| $*_{Pt}$-$H_{Pt}$-$O_bH$ (IM8) + $O_s$ + $CO_{(g)}$ + $CO_{2(g)}$→ $*_{Pt}$-$H_{Pt}$-$O_sH$ (IM9) + $O_b$+ $CO_{(g)}$ + $CO_{2(g)}$ | 0.433 | 0.817 | 0.686 | 1.341 |
| $*_{Pt}$-$H_{Pt}$-$O_sH$ (IM9) + $CO_{(g)}$ + $CO_{2(g)}$ | 0.366 | 0.713 | 0.552 | 0.953 |
| $*_{Pt}$-$H_{Pt}$-$O_sH$ (IM9) + $CO_{(g)}$ + $CO_{2(g)}$ → $2H_{Pt}$ (IM10) + $O_s$+ $CO_{(g)}$ + $CO_{2(g)}$ | 0.556 | 0.880 | 0.881 | 1.450 |
| $2H_{Pt}$ (IM10) + $CO_{(g)}$ + $CO_{2(g)}$ | -0.272 | -0.009 | -0.210 | 0.345 |
| $2H_{Pt}$ (IM10) + $CO_{(g)}$ + $CO_{2(g)}$→ $*_{Pt}$ (IM1) + $*_{Pt}$+ $CO_{(g)}$ + $CO_{2(g)}$ + $H_{2(g)}$ | -0.272 | -0.009 | -0.210 | 0.345 |
| $*_{Pt}$ (IM1) + $*_{Pt}$+ $CO_{(g)}$ + $CO_{2(g)}$ + $H_{2(g)}$ | -0.638 | -0.590 | -0.440 | -0.772 |
| $CO_{Pt}$ (IM2) + $CO_{(g)}$ + $H_2O_{(g)}$→ $(CO,CO)_{Pt}$ (IM11) + $H_2O_{(g)}$ | -0.468 | -0.194 | -0.469 | -0.014 |
| $(CO,CO)_{Pt}$ (IM11) + $O_i$ + $H_2O_{(g)}$ | -0.834 | -0.775 | -0.699 | -1.131 |
| $(CO,CO)_{Pt}$ (IM11) + $O_i$ + $H_2O_{(g)}$→ $(CO,CO_2)_{Pt-Oi}$ (IM12) + $H_2O_{(g)}$ | -0.307 | -0.037 | -0.113 | -0.547 |
| $(CO,CO_2)_{Pt-Oi}$ (IM12) + $H_2O_{(g)}$ | -0.432 | -0.152 | -0.205 | -0.685 |
| $(CO,CO_2)_{Pt-Oi}$ (IM12) + $H_2O_{(g)}$→ $CO_{Pt}$-$V_i$ (IM13) + $H_2O_{(g)}$+ $CO_{2(g)}$ | 0.900 | 1.038 | 2.083 | 0.256 |
| $CO_{Pt}$-$V_i$ (IM13) + $H_2O_{(g)}$+ $CO_{2(g)}$ | -0.466 | -0.633 | 0.824 | -1.026 |
| $CO_{Pt}$-$V_i$ (IM13) + $H_2O_{(g)}$+ $CO_{2(g)}$→ $CO_{Pt}$-$H_2O_i$ (IM14) + $CO_{2(g)}$ | -0.466 | -0.633 | 0.824 | -1.026 |
| $CO_{Pt}$-$H_2O_i$ (IM14) + $*_{Pt}$ + $CO_{2(g)}$ | -0.287 | -0.114 | 0.842 | -0.617 |
| $CO_{Pt}$-$H_2O_i$ (IM14) + $*_{Pt}$ + $CO_{2(g)}$ → $CO_{Pt}$-$H_{Pt}$-$O_iH$ (IM15) + $CO_{2(g)}$ | -0.253 | -0.108 | 0.518 | -0.550 |
| $CO_{Pt}$-$H_{Pt}$-$O_iH$ (IM15) + $CO_{2(g)}$ | -1.114 | -0.896 | -0.886 | -1.338 |
| $CO_{Pt}$-$H_{Pt}$-$O_iH$ (IM15) + $CO_{2(g)}$→ $(CO,H)_{Pt}$-$H_{Pt}$ (IM16) + $O_i$ + $CO_{2(g)}$ | -1.051 | -0.842 | -0.774 | -1.096 |
| $(CO,H)_{Pt}$-$H_{Pt}$ (IM16) + $CO_{2(g)}$ | -1.246 | -1.152 | -1.120 | -1.445 |
| $(CO,H)_{Pt}$-$H_{Pt}$ (IM16) + $CO_{2(g)}$ → $CO_{Pt}$ (IM2) + $*_{Pt}$+ $CO_{2(g)}$ + $H_2(g)$ | -1.246 | -1.152 | -1.120 | -1.445 |
| $CO_{Pt}$ (IM2) + $*_{Pt}$+ $CO_{2(g)}$ + $H_2(g)$ | -1.324 | -1.245 | -1.030 | -2.101 |

(c) 573 K

| **Intermediate or Transition State** | **ΔG (eV)** | | | |
|---|---|---|---|---|
| | **PBE** | **RPBE** | **HSE** | **M06L** |
| $*_{Pt}$ (IM1) + $2CO_{(g)}$ + $H_2O_{(g)}$ | 0.000 | 0.000 | 0.000 | 0.000 |
| $*_{Pt}$ (IM1) + $2CO_{(g)}$ + $H_2O_{(g)}$→ $CO_{Pt}$ (IM2) + $CO_{(g)}$ + $H_2O_{(g)}$ | 0.000 | 0.000 | 0.000 | 0.000 |
| $CO_{Pt}$ (IM2) + $*_{Ti}$ + $CO_{(g)}$ + $H_2O_{(g)}$ | -0.545 | -0.271 | -0.546 | -0.091 |
| $CO_{Pt}$ (IM2) + $*_{Ti}$ + $CO_{(g)}$ + $H_2O_{(g)}$→$CO_{Pt}$-$H_2O_{Ti}$ (IM3) + $CO_{(g)}$ | -0.545 | -0.271 | -0.546 | -0.091 |
| $CO_{Pt}$-$H_2O_{Ti}$ (IM3) + $O_b$ + $CO_{(g)}$ | -0.111 | 0.421 | -0.111 | 0.054 |
| $CO_{Pt}$-$H_2O_{Ti}$ (IM3) + $O_b$ + $CO_{(g)}$→$CO_{Pt}$-$OH_{Ti}$-$O_bH$ (IM4) + $CO_{(g)}$ | 0.251 | 0.917 | 0.272 | 0.570 |
| $CO_{Pt}$-$OH_{Ti}$-$O_bH$ (IM4) + $CO_{(g)}$ | 0.091 | 0.681 | 0.013 | 0.086 |
| $CO_{Pt}$-$OH_{Ti}$-$O_bH$ (IM4) + $CO_{(g)}$ → $COOH_{(Pt-Ti)}$-$O_bH$ (IM5) + $CO_{(g)}$ | 0.729 | 1.500 | 0.721 | 0.869 |
| $COOH_{(Pt-Ti)}$-$O_bH$ (IM5)+ $O_s$+ $CO_{(g)}$ | 0.588 | 1.376 | 0.527 | 0.836 |
| $COOH_{(Pt-Ti)}$-$O_bH$ (IM5)+ $O_s$+ $CO_{(g)}$ →$CO_{2(Pt-Ti)}$-$O_bH$-$O_sH$ (IM6) + $CO_{(g)}$ | 1.183 | 2.072 | 1.275 | 1.928 |
| $CO_{2(Pt-Ti)}$-$O_bH$-$O_sH$ (IM6) + $CO_{(g)}$ | 1.089 | 1.997 | 1.107 | 1.408 |
| $CO_{2(Pt-Ti)}$-$O_bH$-$O_sH$ (IM6) + $CO_{(g)}$ → $*_{Pt}$-$O_bH$-$O_sH$ (IM7) + $*_{Ti}$ + $CO_{(g)}$ + $CO_{2(g)}$ | 1.089 | 1.997 | 1.107 | 1.408 |
| $*_{Pt}$-$O_bH$-$O_sH$ (IM7) + $*_{Pt}$ + $CO_{(g)}$ + $CO_{2(g)}$ | 0.458 | 0.916 | 1.007 | 0.434 |
| $*_{Pt}$-$O_bH$-$O_sH$ (IM7) + $*_{Pt}$ + $CO_{(g)}$ + $CO_{2(g)}$ → $*_{Pt}$-$H_{Pt}$-$O_bH$ (IM8) + $O_s$+ $CO_{(g)}$ + $CO_{2(g)}$ | 0.769 | 1.211 | 1.616 | 1.256 |
| $*_{Pt}$-$H_{Pt}$-$O_bH$ (IM8) + $O_s$+ $CO_{(g)}$ + $CO_{2(g)}$ | -0.183 | 0.152 | -0.037 | 0.047 |
| $*_{Pt}$-$H_{Pt}$-$O_bH$ (IM8) + $O_s$ + $CO_{(g)}$ + $CO_{2(g)}$→ $*_{Pt}$-$H_{Pt}$-$O_sH$ (IM9) + $O_b$+ $CO_{(g)}$ + $CO_{2(g)}$ | 0.494 | 0.878 | 0.747 | 1.402 |
| $*_{Pt}$-$H_{Pt}$-$O_sH$ (IM9) + $CO_{(g)}$ + $CO_{2(g)}$ | 0.401 | 0.748 | 0.587 | 0.988 |
| $*_{Pt}$-$H_{Pt}$-$O_sH$ (IM9) + $CO_{(g)}$ + $CO_{2(g)}$ → $2H_{Pt}$ (IM10) + $O_s$+ $CO_{(g)}$ + $CO_{2(g)}$ | 0.632 | 0.956 | 0.957 | 1.526 |
| $2H_{Pt}$ (IM10) + $CO_{(g)}$ + $CO_{2(g)}$ | -0.224 | 0.039 | -0.162 | 0.393 |
| $2H_{Pt}$ (IM10) + $CO_{(g)}$ + $CO_{2(g)} \rightarrow *_{Pt}$ (IM1) + $*_{Pt}$+ $CO_{(g)}$ + $CO_{2(g)}$ + $H_{2(g)}$ | -0.224 | 0.039 | -0.162 | 0.393 |
| $*_{Pt}$ (IM1) + $*_{Pt}$+ $CO_{(g)}$ + $CO_{2(g)}$ + $H_{2(g)}$ | -0.616 | -0.568 | -0.418 | -0.750 |
| $CO_{Pt}$ (IM2) + $CO_{(g)}$ + $H_2O_{(g)} \rightarrow (CO,CO)_{Pt}$ (IM11) + $H_2O_{(g)}$ | -0.044 | 0.230 | -0.045 | 0.410 |
| $(CO,CO)_{Pt}$ (IM11) + $O_i$ + $H_2O_{(g)}$ | -0.872 | -0.813 | -0.737 | -1.169 |
| $(CO,CO)_{Pt}$ (IM11) + $O_i$ + $H_2O_{(g)} \rightarrow (CO,CO_2)_{Pt\text{-}Oi}$ (IM12) + $H_2O_{(g)}$ | -0.237 | 0.033 | -0.043 | -0.477 |
| $(CO,CO_2)_{Pt\text{-}Oi}$ (IM12) + $H_2O_{(g)}$ | -0.397 | -0.117 | -0.170 | -0.650 |
| $(CO,CO_2)_{Pt\text{-}Oi}$ (IM12) + $H_2O_{(g)} \rightarrow CO_{Pt}\text{-}V_i$ (IM13) + $H_2O_{(g)}$+ $CO_{2(g)}$ | 0.500 | 0.638 | 1.683 | -0.144 |
| $CO_{Pt}\text{-}V_i$ (IM13) + $H_2O_{(g)}$+ $CO_{2(g)}$ | -0.284 | -0.451 | 1.006 | -0.844 |
| $CO_{Pt}\text{-}V_i$ (IM13) + $H_2O_{(g)}$+ $CO_{2(g)} \rightarrow CO_{Pt}\text{-}H_2O_i$ (IM14) + $CO_{2(g)}$ | -0.284 | -0.451 | 1.006 | -0.844 |
| $CO_{Pt}\text{-}H_2O_i$ (IM14) + $*_{Pt}$ + $CO_{2(g)}$ | -0.167 | 0.006 | 0.962 | -0.497 |
| $CO_{Pt}\text{-}H_2O_i$ (IM14) + $*_{Pt}$ + $CO_{2(g)} \rightarrow CO_{Pt}\text{-}H_{Pt}\text{-}O_iH$ (IM15) + $CO_{2(g)}$ | -0.073 | 0.072 | 0.698 | -0.370 |
| $CO_{Pt}\text{-}H_{Pt}\text{-}O_iH$ (IM15) + $CO_{2(g)}$ | -0.945 | -0.727 | -0.717 | -1.169 |
| $CO_{Pt}\text{-}H_{Pt}\text{-}O_iH$ (IM15) + $CO_{2(g)} \rightarrow (CO,H)_{Pt}\text{-}H_{Pt}$ (IM16) + $O_i$ + $CO_{2(g)}$ | -0.855 | -0.646 | -0.578 | -0.900 |
| $(CO,H)_{Pt}\text{-}H_{Pt}$ (IM16) + $CO_{2(g)}$ | -1.090 | -0.996 | -0.964 | -1.289 |
| $(CO,H)_{Pt}\text{-}H_{Pt}$ (IM16) + $CO_{2(g)} \rightarrow CO_{Pt}$ (IM2) + $*_{Pt}$+ $CO_{2(g)}$ + $H_2(g)$ | -1.090 | -0.996 | -0.964 | -1.289 |
| $CO_{Pt}$ (IM2) + $*_{Pt}$+ $CO_{2(g)}$ + $H_2(g)$ | -1.161 | -1.082 | -0.867 | -1.938 |

## I (d) Prior construction for gas molecule corrections and model discrepancy

In order to form a Dirichlet probability density function for correcting the thermodynamics as performed by Walker et al.[5], a range is necessary to set on the individual gas molecule corrections. In this study the range $-0.6 \leq \zeta \leq 0.6\ (eV)$ has been chosen to weakly inform the initial prior, which is further constrained using the first experimental data as previously described.

The prior uncertainty for the standard deviations, $\sigma$, corresponding to the discrepancy model in Equation 4 of the main paper is given by an inverse gamma probability density function with parameters $\alpha$, $\beta$ as listed in Supplementary Table 4.

Supplementary Table 4. Discrepancy model standard deviations inverse gamma priors. These hyperparameters are given a prior uncertainty, which is tuned during the first Bayesian inverse problem.

| $TOF\ (s^{-1})$ | | $\alpha_{CO}$ | | $\alpha_{H_2O}$ | | $\alpha_{CO_2}$ | | $\alpha_{H_2}$ | | $E_{act}\ (eV)$ | |
|---|---|---|---|---|---|---|---|---|---|---|---|
| $\alpha$ | $\beta$ | $\alpha$ | $\beta$ | $\alpha$ | $\beta$ | $\alpha$ | $\beta$ | $\alpha$ | $\beta$ | $\alpha$ | $\beta$ |
| 3 | 4 | 3 | 0.4 | 3 | 0.4 | 3 | 0.4 | 3 | 0.4 | 3 | 0.8 |

## II. Experimental data

Supplementary Table 5. Experimental conditions for D1[14], D2[15] and D3[16] and their respective quantities of interest.

| | **D1** | **D2** | **D3** |
|---|---|---|---|
| | T=503K | T=523K | T=573K |
| log10(TOF) (1/s) | -0.89 | -0.89 | -0.22 |
| APP (eV) | 0.58 | 0.47 | 0.61 |
| CO (atm) | 0.07 | 0.03 | 0.10 |
| $CO_2$ (atm) | 0.09 | 0.06 | 0.10 |
| $H_2$ (atm) | 0.37 | 0.20 | 0.40 |
| $H_2O$ (atm) | 0.22 | 0.10 | 0.20 |
| $\alpha_{CO}$ | 0.30 | 0.50 | 0.30 |
| $\alpha_{CO2}$ | 0.00 | 0.00 | 0.00 |
| $\alpha_{H2}$ | -0.70 | -0.70 | -0.67 |
| $\alpha_{H2O}$ | 0.68 | 1.00 | 0.85 |

## III. Model validation via posterior predictive check

Due to the inconclusive model discrimination based on evidence values of Table 2 of the main paper, a test is conducted to select the site model, edge or corner, that best agrees with both calibration and validation data. The test is a chi-squared test of the squared Mahalanobis distance[17] for all three experimental data points. The key principle of the Mahalanobis distance is that a multidimensional experimental data point may fall outside of the assumed multidimensional Gaussian uncertainty even though the marginal uncertainties capture the observational data. In this study, there are 6 dimensions, one for TOF ($s^{-1}$), four reaction orders, and the apparent activation barrier (eV).

The square of the Mahalanobis distance is

$$(\boldsymbol{d}-\boldsymbol{\mu})'\boldsymbol{\Sigma}^{-1}(\boldsymbol{d}-\boldsymbol{\mu}) \tag{S7}$$

where $\boldsymbol{\mu}$ is a 6x1 vector of predictive mean values, $\boldsymbol{\Sigma}$ is a 6x6 predictive covariance matrix, and $\boldsymbol{d}$ is a 6x1 vector of experimental data. The predictive distribution includes the uncertainty due to model error. The six quantities are turnover frequency ($s^{-1}$), four reaction orders, and apparent activation barrier (eV). When the square of the Mahalanobis distance passes the following test, the experimental data is considered to *not* be an outlier and thus, a plausible outcome of the model in question.

$$(\boldsymbol{d}-\boldsymbol{\mu})'\boldsymbol{\Sigma}^{-1}(\boldsymbol{d}-\boldsymbol{\mu}) \leq \chi^2_{0.05} \quad \text{(S8)}$$

For six-dimensional space, $\chi^2_{0.05} = 12.592$.

## IV. Terrace active site model

In this section, we present a more detailed description of our Pt(111) WGS model. All DFT energies and a microkinetic model are summarized below following the reaction mechanisms studied by Grabow et al.[18] for Pt(111). This study uses the Perdew-Burke-Ernzerhof (PBE)[7] functional for all DFT calculations. All calculations have been performed with the Vienna Ab initio Simulation Package (VASP).[19,20] Climbing image nudged elastic band and dimer methods are used to locate first order saddle points, i.e., the transition states.[21-23] Transition state theory and collision theory are used to calculate elementary rate constants. Transition state theory is expressed as

$$k = \frac{k_B T}{h}\exp\left(-\frac{\Delta G^{\ddagger}}{k_B T}\right) \quad \text{(S9)}$$

where $k$ is the rate constant, $k_B$ is Boltzmann's constant, $h$ is Planck's constant, $T$ is temperature, and $\Delta G^{\ddagger}$ is the free energy activation barrier. Lateral interactions due to two abundant surface intermediates, CO and H, are applied to all intermediates and transition states (see below). A 3x4x4 slab model of 48 Pt atoms is used for all DFT calculations. The top two layers are relaxed and the bottom two layers are fixed in the calculations. Calculated lattice constants are $a = 8.4342$ Å, $b = 9.7389$ Å, $c = 22.0000$ Å. A kinetic energy cutoff of 400 eV and a Gaussian smearing of 0.1 eV have been used in the DFT calculations.

Equations (S10)-(S24) summarize the elementary reaction steps considered in the microkinetic model used to solve for turnover frequency (TOF $s^{-1}$), reaction orders, and apparent activation barrier (eV).

$$* + CO \leftrightarrow CO^* \tag{S10}$$

$$* + H_2O \leftrightarrow H_2O^* \tag{S11}$$

$$* + H_2O^* \leftrightarrow H^* + OH^* \tag{S12}$$

$$* + OH^* \leftrightarrow H^* + O^* \tag{S13}$$

$$OH^* + OH^* \leftrightarrow H_2O^* + O^* \tag{S14}$$

$$CO^* + O^* \leftrightarrow CO_2^* + * \tag{S15}$$

$$* + COOH^* \leftrightarrow CO_2^* + H^* \tag{S16}$$

$$COOH^* + O^* \leftrightarrow CO_2^* + OH^* \tag{S17}$$

$$COOH^* + OH^* \leftrightarrow CO_2^* + H_2O^* \tag{S18}$$

$$HCOO^* + * \leftrightarrow CO_2^* + H^* \tag{S19}$$

$$HCOO^* + O^* \leftrightarrow CO_2^* + OH^* \tag{S20}$$

$$HCOO^* + OH^* \leftrightarrow CO_2^* + H_2O^* \quad \text{(S21)}$$

$$2H^* \leftrightarrow H_2 + 2* \quad \text{(S22)}$$

$$CO_2^* \leftrightarrow CO_2 + * \quad \text{(S23)}$$

$$HCO^* + * \leftrightarrow CO^* + H^* \quad \text{(S24)}$$

Asterisks, *, refers to an adsorbed species or a vacant surface site if the asterisk is unaccompanied by a chemical species. Species without an asterisk are present in the gas phase. Supplementary Table 6 summarize the DFT results for the water-gas shift (WGS) on Pt(111).

Supplementary Table 6. Zero-point corrected activation barriers (eV) and reaction energies (eV) for elementary steps of the water-gas shift (WGS) reaction on Pt(111).

| Reaction | Activation barrier (eV) | Reaction energy (eV) |
|---|---|---|
| $CO +^* \leftrightarrow CO^*$ | 0.0 | -1.76 |
| $H_2O +^* \leftrightarrow H_2O^*$ | 0.0 | -0.14 |
| $H_2O^* +^* \leftrightarrow H^* + OH^*$ | 0.77 | 0.42 |
| $OH^* +^* \leftrightarrow H^* + O^*$ | 0.89 | -0.15 |
| $OH^* + OH^* \leftrightarrow H_2O^* + O^*$ | -0.04 | -0.58 |
| $CO^* + O^* \leftrightarrow CO_2^* +^*$ | 1.06 | -0.18 |
| $CO^* + OH^* \leftrightarrow COOH^* +^*$ | 0.43 | -0.11 |
| $COOH^* +^* \leftrightarrow CO_2^* + H^*$ | 0.60 | -0.22 |
| $COOH^* + O^* \leftrightarrow CO_2^* + OH^*$ | 0.23 | -0.07 |
| $COOH^* + OH^* \leftrightarrow CO_2^* + H_2O^*$ | 0.0 | -0.67 |
| $HCOO^* +^* \leftrightarrow CO_2^* + H^*$ | 0.99 | -0.57 |
| $HCOO^* + O^* \leftrightarrow CO_2^* + OH^*$ | 1.27 | -0.45 |
| $HCOO^* + OH^* \leftrightarrow CO_2^* + H_2O^*$ | 0.93 | -1.01 |
| $2H^* \leftrightarrow H_2 + 2^*$ | 0.0 | 0.96 |
| $CO_2^* \leftrightarrow CO_2 +^*$ | 0.0 | -0.05 |
| $HCO^* +^* \leftrightarrow CO^* + H^*$ | 0.09 | -1.15 |

## Lateral interaction model

The lateral interaction effects on surface intermediates and transition states due to the most abundant surface intermediates, CO and H, are considered. Without considering lateral interaction effects, the Pt(111) WGS microkinetic model becomes poisoned by CO (greater than 99.9% CO surface coverage) and the activity is approximately 9 orders of magnitude less than with lateral interactions. Due to the nature of the catalyst model for the corner and edge active sites those models incorporate lateral interaction effects already.

We begin this section with an explanation of lateral interaction effects for Pt(111) including a comparison of the relative free energy pathways with and without lateral interaction effects. Next, Campbell's degree of rate control (DRC)[13,24-27] is explained and the sum of the DRC for all the elementary steps is shown to be one for Pt(111) with lateral interaction effects. Finally, a comparison of the results on the WGS over Pt(111) by Stamatakis et al.[28] is conducted.

Supplementary Figures 1 and 2 are examples of the lateral interaction effects due to CO and H. In fact, every intermediate and transition state includes linear lateral interaction effects due to CO and H such as those show in Supplementary Figures 1 and 2. We note that Vlachos et al. used a similar lateral interaction model.[29] The lateral interaction effect for any species $i$, $\Delta E_i^{lat}$, is

$$\Delta E_i^{lat} = m_{i,H}\theta_H + m_{i,CO}\theta_{CO} \tag{S25}$$

where $\theta_H$ and $\theta_{CO}$ are the dimensionless surface coverage of hydrogen and CO. The slope of the linear relationship between the surface coverage and lateral interaction effect is $m_{i,j}$. The slope is obtained through a linear fit of differential adsorption energies computed by DFT. The structures for the DFT calculations were obtained by first determining the most stable configuration of CO and H themselves on the unit cell surface at various coverages. The most stable configuration was then used for all other species. To accelerate the calculation of the adsorption energy of any surface species at various CO and hydrogen coverage, we relaxed in these DFT calculations only the CO and H coordinates, i.e., the other atoms were fixed in their original optimized position.

(a)

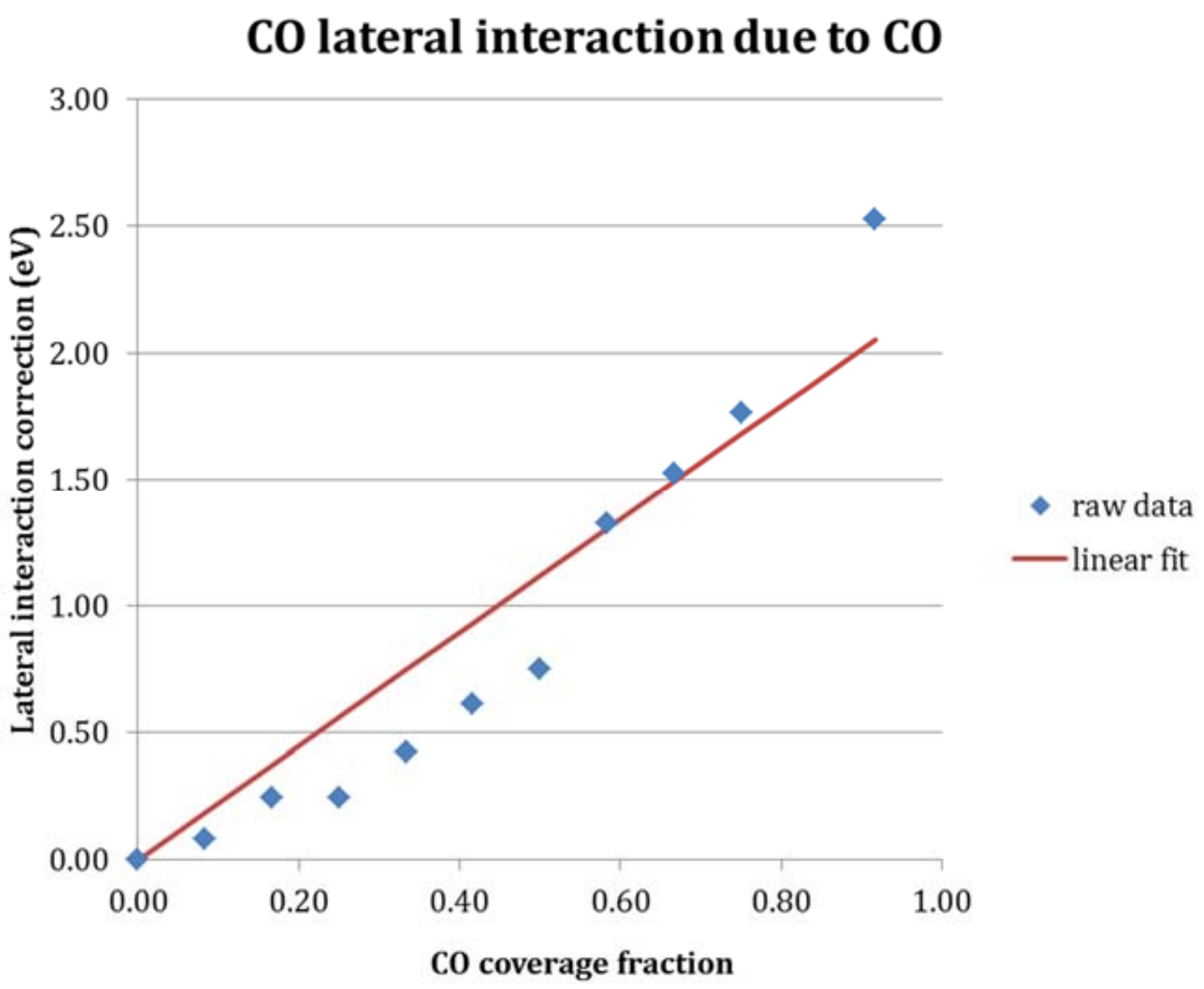

(b)

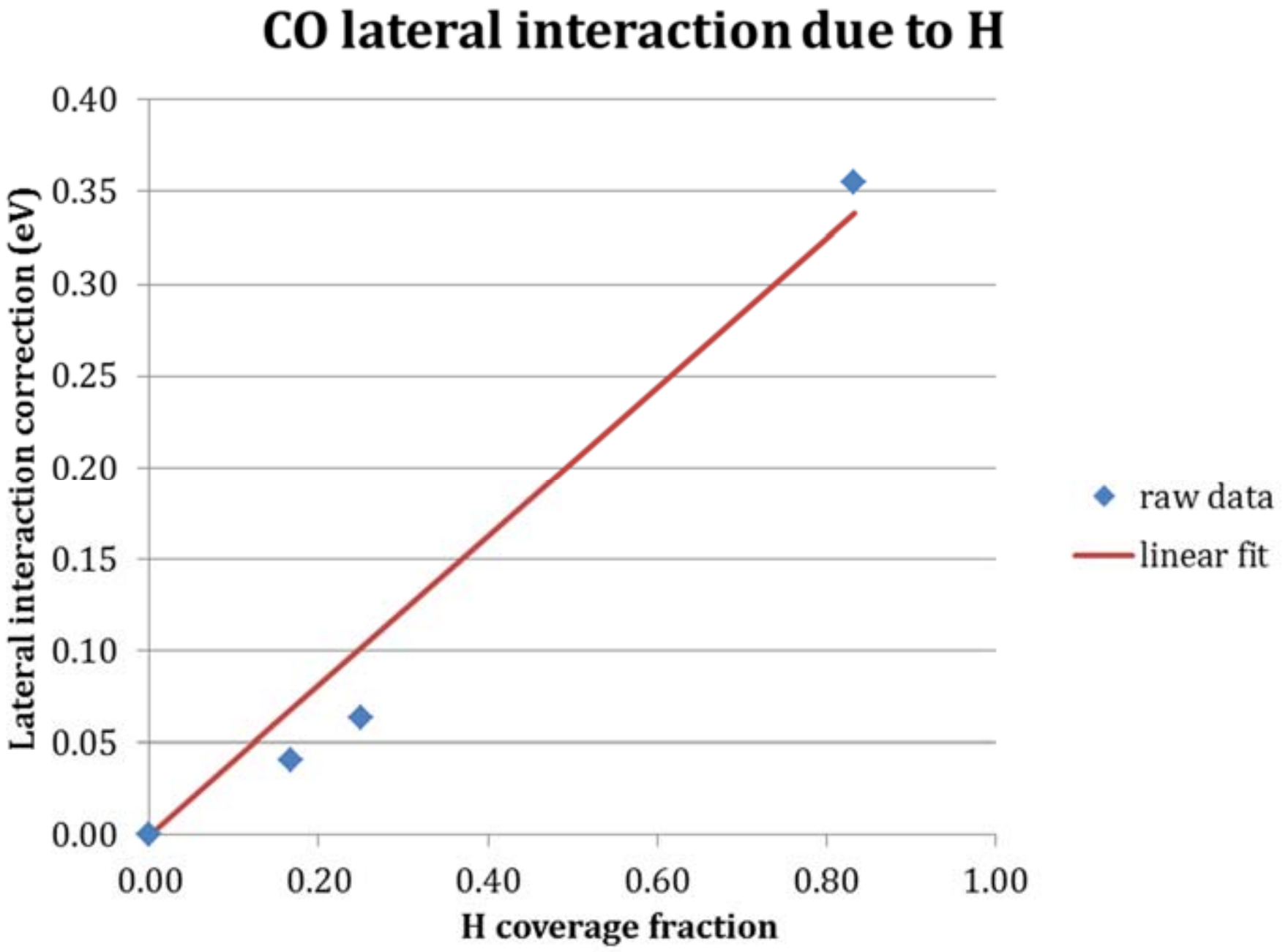

Supplementary Figure 1. Example lateral interaction effects (eV) on the CO surface intermediate. A linear functional dependence on CO and H dimensionless surface coverage fraction for lateral interaction effect (eV) is shown for the example of the CO surface intermediate. All surface intermediates and transition states have such a linear functional dependence. (a) CO lateral interaction due to CO. (b) CO lateral interaction due to H.

(a)

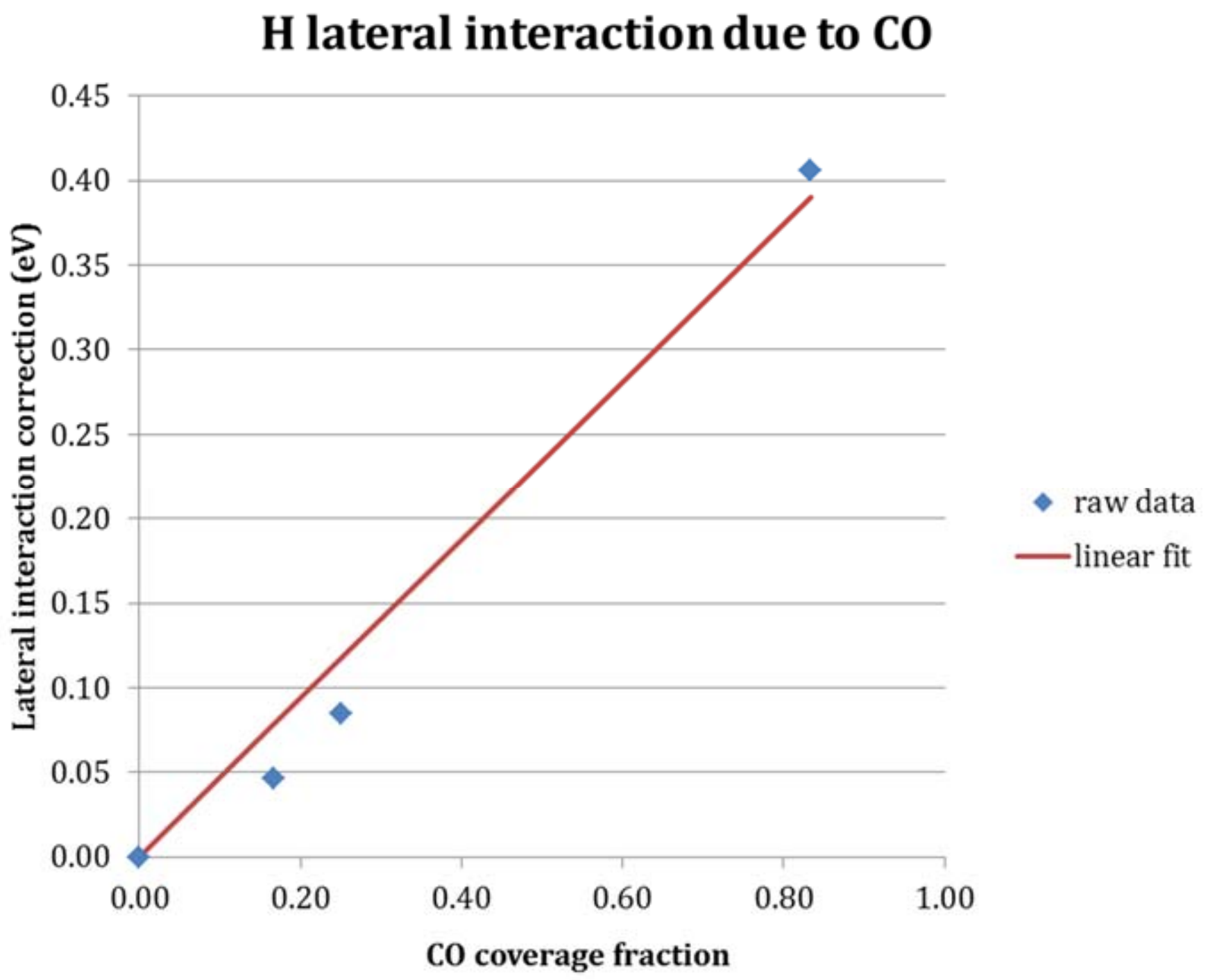

(b)

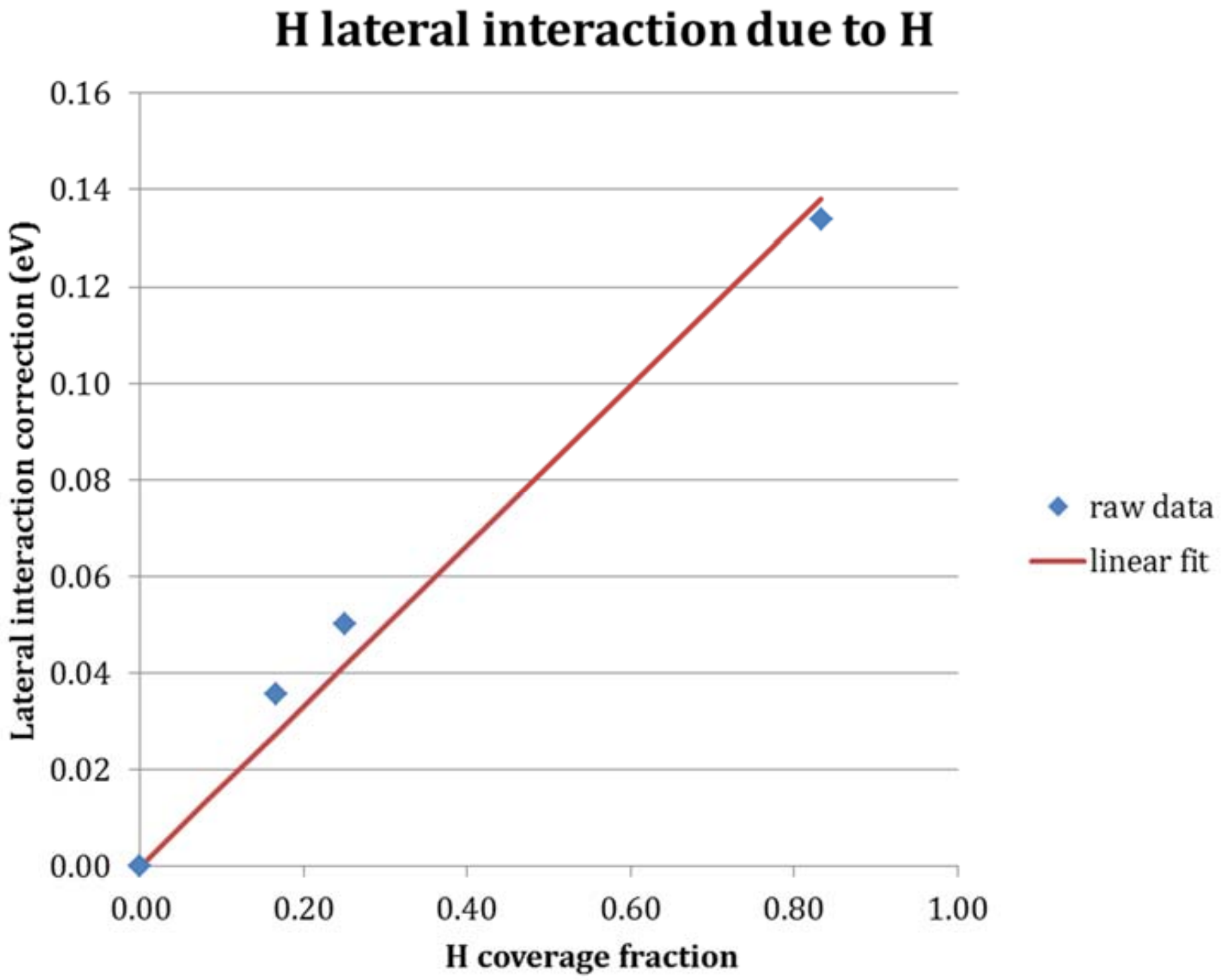

Supplementary Figure 2**.** Example lateral interaction effects (eV) on H surface intermediate. (a) H lateral interaction due to CO. (b) H lateral interaction due to H.

As stated above, all intermediates and transition states are affected by the CO and H surface coverage. Supplementary Table 7 lists the lateral interaction parameters of each surface and transition state species due to a varying CO and H coverage.

Supplementary Table 7. Lateral interaction parameters of CO and H on each surface and transition state species programmed into our Pt(111) WGS microkinetic model.

| Species *i* | $m_{i,CO}$ | $m_{i,H}$ |
|---|---|---|
| $H^*$ | 0.47 | 0.17 |
| $O^*$ | 2.03 | 0.90 |
| $OH^*$ | -1.13 | 0.42 |
| $H_2O^*$ | -0.88 | 0.16 |
| $CO^*$ | 2.52 | 0.41 |
| $CO_2^*$ | -1.08 | 0.01 |
| $COOH^*$ | 1.86 | 0.33 |
| $HCOO^*$ | 2.41 | 0.72 |
| $HCO^*$ | 1.63 | 0.38 |
| $H_2O^* +^* \leftrightarrow H^* + OH^*$ | 3.47 | 0.36 |
| $OH^* +^* \leftrightarrow H^* + O^*$ | 4.16 | 0.42 |
| $OH^* + OH^* \leftrightarrow H_2O^* + O^*$ | -1.18 | 0.42 |
| $CO^* + O^* \leftrightarrow CO_2^* +^*$ | -1.69 | -0.23 |
| $CO^* + OH^* \leftrightarrow COOH^* +^*$ | 2.82 | 0.16 |
| $COOH^* +^* \leftrightarrow CO_2^* + H^*$ | 0.75 | 0.01 |
| $COOH^* + O^* \leftrightarrow CO_2^* + OH^*$ | 1.20 | 0.04 |
| $COOH^* + OH^* \leftrightarrow CO_2^* + H_2O^*$ | 2.23 | 0.37 |
| $HCOO^* +^* \leftrightarrow CO_2^* + H^*$ | 1.43 | -2.62 |
| $HCOO^* + O^* \leftrightarrow CO_2^* + OH^*$ | 0.00 | -0.54 |
| $HCOO^{*} + OH^{*} \leftrightarrow CO_2^{*} + H_2O^{*}$ | 1.38 | 0.27 |
| $HCO^{*} +^{*} \leftrightarrow CO^{*} + H^{*}$ | 1.56 | 0.18 |

The free energy paths with no lateral interactions and with lateral interactions are compared for the dominant (highest activity) reaction pathway in Supplementary Figure 3. The free energy path with lateral interactions represent the steady state pathway achieved by the microkinetic model, i.e., at T=523 (K), $P_{CO} = 0.03$ (atm), $P_{H_2O} = 0.1$, $P_{CO_2} = 0.06$, $P_{H_2} = 0.2$.

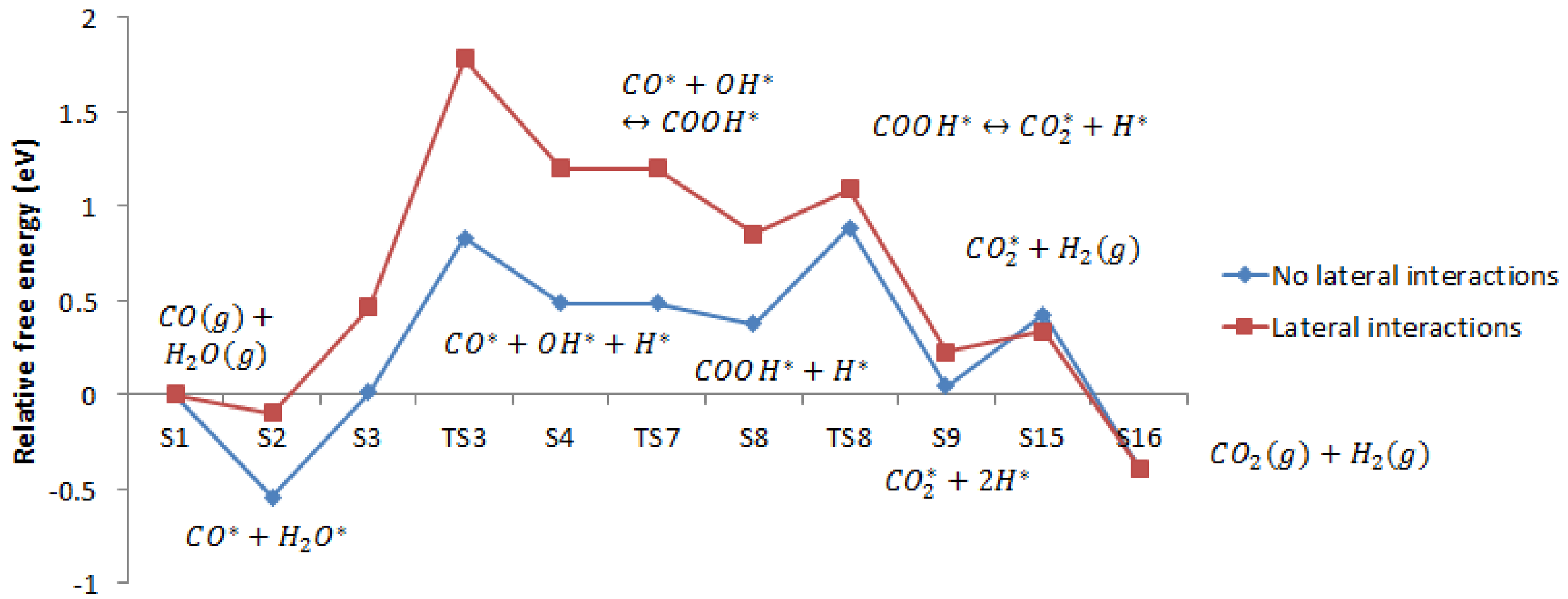

Supplementary Figure 3. Comparison of relative free energy paths for no lateral interactions and with linear lateral interactions. The free energy paths represent the steady state achieved by the microkinetic model. The overall reaction thermodynamics are not affected by the inclusion of lateral interactions. The dominant (most active) reaction pathway is displayed. T=523 (K), $P_{CO} = 0.03$ (atm), $P_{H_2O} = 0.1$, $P_{CO_2} = 0.06$, $P_{H_2} = 0.2$.

Next, we introduce the use of Campbell's DRC and apply it to our Pt(111) model.[13,24-27] Campbell's DRC may be applied to any microkinetic model. Sutton et al.[30] have applied a similar sensitivity measure to Campbell's DRC under uncertainty. Later in this work, we analyze the uncertainty in DRC for the edge active site model. The DRC for each elementary step in the WGS on Pt(111) is listed in Supplementary Table 8. The formula for the DRC is,[13,24-27]

$$X_{RC,i} = \left( \frac{\partial lnTOF}{\partial \frac{-G_i^{TS}}{k_B T}} \right)_{G_{j \neq i}^{TS}, G_m} \quad \text{(S26)}$$

$X_{RC,i}$ is the degree of rate control for transition state $i$. $G_i^{TS}$ is the relative free energy of transition state structure $i$. The temperature and all transition state relative free energies $j$ (and intermediates relative free energies $m$) are held constant. It should be noted that a property of the DRC is that the sum of all DRC for a microkinetic model are one,[27,31]

$$\Sigma_i X_{RC,i} = 1 \quad \text{(S27)}$$

This property holds true with the lateral interactions used is this work. The last row of Supplementary Table 8 shows the DRC summing to one.

Supplementary Table 8**.** Campbell's degree of rate control (DRC) for each elementary step of the WGS on Pt(111). These DRC correspond to the lateral interaction free energy pathway in Supplementary Figure 3. The last row shows the sum of all DRC for each elementary reaction step to be one, which is a verification for the implementation of the lateral interaction effects. T=523 (K), $P_{CO} = 0.03$ (atm), $P_{H_2O} = 0.1$, $P_{CO_2} = 0.06$, $P_{H_2} = 0.2$.

| Elementary reaction step | Degree of rate control |
|---|---|
| $* + CO \leftrightarrow CO^*$ | -3.5×10$^{-12}$ |
| $* + H_2O \leftrightarrow H_2O^*$ | -5.5×10$^{-12}$ |
| $* + H_2O \leftrightarrow H^* + OH^*$ | 0.9995 |
| $* + OH \leftrightarrow H^* + O^*$ | -6.4×10$^{-12}$ |
| $OH^* + OH^* \leftrightarrow H_2O^* + O^*$ | 0.0 |
| $CO^* + O^* \leftrightarrow CO_2^* +^*$ | 2.5×10$^{-12}$ |
| $CO^* + OH^* \leftrightarrow COOH^* +^*$ | 3.9×10$^{-4}$ |
| $COOH^* +^* \leftrightarrow CO_2^* + H^*$ | 1.3×10$^{-6}$ |
| $COOH^* + O^* \leftrightarrow CO_2^* + OH^*$ | 1.1×10$^{-11}$ |
| $COOH^* + OH^* \leftrightarrow CO_2^* + H_2O^*$ | 0.0 |
| $HCOO^* +^* \leftrightarrow CO_2^* + H^*$ | 1.1×10$^{-11}$ |
| $HCOO^* + O^* \leftrightarrow CO_2^* + OH^*$ | -5.5×10$^{-12}$ |
| $HCOO^* + OH^* \leftrightarrow CO_2^* + H_2O^*$ | -2.0×10$^{-12}$ |
| $2H^* \leftrightarrow H_2 + 2^*$ | -6.5×10$^{-12}$ |
| $CO_2^* \leftrightarrow CO_2 +^*$ | 3.5×10$^{-12}$ |
| $HCO^{*}+^{*}\leftrightarrow CO^{*}+H^{*}$ | $8.6\times10^{-12}$ |
| Sum of all degrees of rate control | **0.99990** |

Next, Stamatakis et al.[28] have found, using graph-theoretical kinetic Monte Carlo (kMC) simulations, a higher TOF ($s^{-1}$) and lower apparent activation barrier for the WGS over Pt(111) than observed in our microkinetic model. For comparison, Supplementary Table 9 lists both our and Stamatakis et al.'s pre-exponential factors, A, for rate constants and equilibrium constants. Interestingly, using our zero-point corrected DFT energies and mean-field microkinetic model with the pre-exponential factors and lateral interactions from Stamatakis et al.,[31] we obtain at 650 K a TOF which is on the same order of magnitude as Stamatakis et al.[31] Therefore, we conclude that the cause for the different TOFs is in the pre-exponential factors and lateral interactions and not in the use of a mean-field microkinetic model versus a kMC simulation (or the DFT energies). We note that we directly computed lateral interaction effects of all surface intermediates *and* transition states while Stamatakis et al. focused on including lateral interaction effects only for some surface species. Since lateral interactions are generally repulsive for most metal catalysts, not including lateral interactions for transition states leads to (unphysically) large elementary rate constants, possibly explaining the difference between Stamatakis et al. and our results.

Supplementary Table 9. Preexponential factors computed at T=650 K computed by Stamatakis et al. and those computed in this study.

| Elementary step | $A_{fwd}$ ($s^{-1}$) Stamatakis et al.[31] | $A_{fwd}/A_{bwd}$ Stamatakis et al.[31] | $A_{fwd}$ ($s^{-1}$) Walker et al. | $A_{fwd}/A_{bwd}$ Walker et al. |
|---|---|---|---|---|
| $* + CO \leftrightarrow CO^*$ | $3.41\times10^{5}$ | $3.43\times10^{-9}$ | $1.34\times10^{8}$ | $2.85\times10^{-8}$ |
| $* + H_2O \leftrightarrow H_2O^*$ | $7.20\times10^{5}$ | $1.69\times10^{-7}$ | $1.67\times10^{8}$ | $1.47\times10^{-6}$ |
| $* + H_2O \leftrightarrow H^* + OH^*$ | $4.48\times10^{12}$ | 6.39 | $4.53\times10^{12}$ | $2.90\times10^{-1}$ |
| $* + OH \leftrightarrow H^* + O^*$ | $2.36\times10^{13}$ | 19.8 | $3.49\times10^{13}$ | $5.20\times10^{-1}$ |
| $OH^* + OH^* \leftrightarrow H_2O^* + O^*$ | $3.09\times10^{11}$ | 3.09 | $1.27\times10^{13}$ | 1.27 |
| $CO^* + O^* \leftrightarrow CO_2^* + ^*$ | $1.17\times10^{12}$† | $1.77\times10^{7}$† | $4.42\times10^{13}$ | 8.25 |
| $CO^* + OH^* \leftrightarrow COOH^* + ^*$ | $4.58\times10^{11}$ | $3.90\times10^{-2}$ | $2.73\times10^{13}$ | $8.91\times10^{-1}$ |
| $COOH^* + ^* \leftrightarrow CO_2^* + H^*$ | $5.28\times10^{14}$ | $8.96\times10^{9}$ | $1.31\times10^{14}$ | $1.31\times10^{1}$ |
| $COOH^* + O^* \leftrightarrow CO_2^* + OH^*$ | $6.93\times10^{11}$ | $4.53\times10^{8}$ | $1.88\times10^{13}$ | 8.40 |
| $COOH^* + OH^* \leftrightarrow CO_2^* + H_2O^*$ | $1.04\times10^{13}$ | $1.40\times10^{9}$ | $5.55\times10^{12}$ | 1.76 |
| $HCOO^* + ^* \leftrightarrow CO_2^* + H^*$ | $5.97\times10^{13}$ | $1.54\times10^{10}$ | $4.29\times10^{13}$ | 3.63 |
| $HCOO^* + O^* \leftrightarrow CO_2^* + OH^*$ | $1.19\times10^{12}$ | $7.76\times10^{8}$ | $1.35\times10^{13}$ | $4.19\times10^{-1}$ |
| $HCOO^* + OH^* \leftrightarrow CO_2^* + H_2O^*$ | $6.30\times10^{12}$ | $2.40\times10^{9}$ | $7.61\times10^{13}$ | 5.19 |
| $2H^* \leftrightarrow H_2 + 2^*$ | $6.17\times10^{12}$ | $2.15\times10^{5}$ | $7.76\times10^{14}$ | $1.55\times10^{6}$ |
| $CO_2^* \leftrightarrow CO_2 + ^*$ | $1.17\times10^{12}$† | $1.77\times10^{7}$† | $6.44\times10^{16}$ | $6.04\times10^{8}$ |
| $HCO^* + ^* \leftrightarrow CO^* + H^*$ | $8.42\times10^{12}$ | $4.26\times10^{1}$ | $1.35\times10^{13}$ | $4.70\times10^{-1}$ |